\documentclass[a4paper,useAMS,usenatbib]{mnras}
\usepackage{color}
\def\lsim{\lower.5ex\hbox{$\; \buildrel < \over \sim \;$}}
\def\gsim{\lower.5ex\hbox{$\; \buildrel > \over \sim \;$}}
\def\be{\begin{equation}}
\def\ee{\end{equation}}

\def\bc{\begin{center}}
\def\ec{\end{center}}
\def\eg{{e.g.}}
\def\etal{{\em et al.}}
\def\ie{{\rm i.e.}}
\def\ep{{{\rm e}-{\rm p}}}

\def\tmn{T^{\mu \nu}}

\def\rg{r_{\rm g}}
\def\rci{r_{\rm ci}}
\def\rjci{r_{\rm jci}}
\def\rcm{r_{\rm cm}}
\def\rjcm{r_{\rm jcm}}
\def\rco{r_{\rm co}}
\def\rjco{r_{\rm jco}}
\def\rsh{r_{\rm sh}}
\def\rmdot{R_{\dot m}}
\def\eni{{\dot{\cal M}}_{\rm ci}}
\def\eno{{\dot{\cal M}}_{\rm co}}
\def\hp{$h_p^2$}
\def\rj{\rm j}
\def\cm{\rm c}
\def\ri{\rm i}
\def\rb{{\rm b}}
\def\vjinf{v_{\rm j \infty}}
\def\rs{r_{\rm s}}
\newcommand{\msol}{\mbox{M$_{\odot}$}}
\newcommand{\mbh}{\mbox{$M_{\rm BH}$}}

\usepackage{graphicx}
\usepackage{mathrsfs}
\usepackage{amsmath}
\title[Bipolar jets around Kerr black holes]
{Estimation of bipolar jets from accretion discs around Kerr black holes}
\author[Kumar \& Chattopadhyay]
{Rajiv Kumar$^{1}$, Indranil Chattopadhyay$^{2}$\thanks{E-mail: kumar@xmu.edu.cn (RK);
indra@aries.res.in (IC)}\\
$^{1}$Department of Astronomy, Xiamen University, Xiamen, Fujian 361005, China\\
$^{2}$Aryabhatta Research Institute of Observational Sciences (ARIES), Manora Peak, Nainital-263002, India\\
}
\begin{document}
\date{}
\maketitle
\label{firstpage}

\begin{abstract}
We analyse flows around a rotating black hole and obtain self-consistent
accretion-ejection solutions in full general relativistic prescription.
Entire energy-angular momentum parameter space is investigated in the advective
regime to obtain shocked and shock-free accretion solutions.
Jet equations of motion are solved along the von-Zeipel surfaces computed from the post-shock disc,
simultaneously with the equations of accretion disc along the equatorial plane. 
For a given spin parameter, the mass outflow rate increases as the shock moves closer to the black hole,
but eventually decreases, maximizing at some intermediate value of shock location.
Interestingly, we obtain all types of possible jet solutions,
for example, steady shock solution with 
multiple critical points, bound solution with two critical points and smooth solution with
single critical point.
Multiple critical points may exist in jet solution for spin parameter $a_s\ge 0.5$.
The jet terminal speed generally increases
if the accretion shock forms closer to the horizon and is
higher for corotating black hole than the counter-rotating and the non-rotating one.
Quantitatively speaking, shocks in jet
may form for spin parameter $a_s>0.6$ and jet shocks range between $6r_g$ and $130r_g$ above the equatorial plane,
while the jet terminal speed $v_{{\rm j}\infty} > 0.35{\cm}$ if Bernoulli parameter ${\cal E}\geq1.01$ for $a_s>0.99$.
\end{abstract}


\section {Introduction}
\label{sec:intro}
Accretion powers many astrophysical objects, like active galactic nuclei (AGNs) and microquasars.
Extreme luminosities of AGNs ($L\sim 10^{42-48}$ erg s$^{-1}$) and micro quasars
($L\sim 10^{36-39}$erg s$^{-1}$) are thought to arise due to accretion of matter and energy 
on to compact objects like black holes (BHs).
AGNs are supposed to harbour supermassive BHs ($\sim 10^{6-9}\msol$, where $\msol$ is the solar
mass) and microquasars are supposed to harbour stellar mass BHs ($\sim 10 \msol$).
In addition,
jets or transonic, relativistic, collimated outflows are quite common and are associated with most of these objects
\citep{mrcpl92,mr94,f98,jbl99,detal12}. Since BHs do not have atmosphere so jets
have to originate from the accreting matter. Although the exact mechanism by which these jets are launched
is still a topic of active research, some salient properties of jets have been established from
observations.
\citet{jbl99} observed that the jet from AGN M87 has to originate from a region $<100 \rs$
($\rs$ is the radius of
non-rotating BH or Schwarzschild radius), while later interpretation of observational data
reduced the estimate of the jet base even further \citep{detal12}.
Such direct estimate of the jet base is yet to be obtained for microquasars. However, since the time-scales
of microquasars and AGNs can be scaled by the central mass, so one can conclude that the
basic physics of AGNs and microquasars are quite similar \citep{mkkf06}.
Moreover, similarities of jet-disc connection were observed in AGNs and microquasars
\citep{marscheretal02}.
This indicates that the jet base for microquasars should also be quite close to the central BH,
as was observed for AGN.

Microquasars undergo a regular luminosity and spectral state changes
within a time-scale of few months. There are two canonical spectral states: the low-hard state or LHS --- low luminosity but radiative
power maximizes in the high-energy power law part; and high-soft state or HSS ---
luminous and power maximizes in the low-energy thermal part of the spectra.
It has been observed that many microquasars in outburst states make 
transition from LHS to HSS through a series of intermediate states, and this cycle repeats in some sort of hysteresis. In the hardness-intensity space this curve looks loosely like a `Q' \citep{fbg04}.  
The most interesting aspect of this property of microquasars is that the jet observed also varies
with this cycle \citep{gfp03}. During HSS there is no jet, while weak quasi-steady jets start to form in LHS.
As the microquasar moves to intermediate hard states, the jet strength increases \citep{rsfp10}. And finally
very strong jets are ejected while the microquasar makes a transition from hard intermediate to soft intermediate states. Eventually, the microquasar enters into HSS, which is luminous but jet is not detected.
The important point these observations have established is that the jet states are deeply linked
with the accretion states, i. e., accretion discs are responsible for the jet generation.
And by virtue of the  similarity between AGNs and microquasars \citep{mkkf06}, one can conclude that even
for AGNs,
the accretion disc is responsible for launching the jet.
  
  It was understood quite early that the accreting matter needs to be rotating because radial accretion would be `too
fast' and would not have the time to generate such high luminosities. The first viable accretion disc model
was proposed by \citet{ss73}, where the disc material possess Keplerian angular momentum, negligible advection
and are optically thick, which quickly radiates the dissipated viscous energy into radiation.The extension to
general relativity was done in the same year by \citet{nt73}. 
Inspite of the obvious theoretical shortcomings such as the simplistic manner with which the inner edge of the disc or the pressure term
was handled, even then, the
Keplerian disc could still explain the thermal, modified
blackbody part of the spectra. Therefore, one may conclude that the HSS is a state dominated
by Keplerian disc. The power-law part of the spectra is not generated by Keplerian disc,
and search for the component of the disc that generates the hard power-law tail in the spectra
launched many competing accretion disc models. Among various disc models, thick disc \citep{pw80} and
advection-dominated accretion flows \citep{ny94,nkh97} are notable. Both Keplerian disc and thick disc models
are rotation dominated with negligible advection and therefore wholly subsonic.
Advection-dominated accretion flows do not suffer from such
limitations, but the boundary conditions were such that the solution is generally subsonic,
and becomes transonic only close to the horizon. Simultaneous to this development, investigations in the advective regime showed some
interesting aspects, like multiple sonic points \citep{lt80}, as well as existence of
steady shocks \citep{f87,c89}. Infact, \citet{lgy99} showed that advection-dominated flows are indeed a subset of general advective accretion solutions. Moreover, it was clear from numerical simulations
that post-shock disc (PSD), due to extra thermal energy gained by shock dissipation can deflect
a fraction of accreting matter along the rotation axis of the disc to produce precursor of jets
\citep{mlc94,msc96,mrc96,lmc98,dcnm14,lckhr16}.
Semi-analytical versions of these studies have been extended into the dissipative
regime for flows described by fixed $\Gamma$ ($=c_p/c_v$ adiabatic index) equation of state
\citep[EoS;][]{bdl08,kc13,kcm14,adn15},
as well as variable $\Gamma$ EoS \citep{kscc13,kc14}.
Since shock in accretion is formed close to the central object ($\rsh \lsim$few$\times 10 \rs$)
and PSD is the base of the jet, therefore, shocked advective discs satisfy one observational criterion,
i. e., the entire accretion discs do not generate jet, only the inner region forms jet.

Existence of the dominant power-law photons in the LHS and intermediate states necessitates existence of
a Comptonizing corona in addition to the Keplerian component of the disc \citep{st80}. Most of the modern accretion disc models differ in the source and location of this corona. For advective shocked
disc, the PSD is the Comptonizing corona. In a model solution \citet{ct95} considered a disc composed 
of Keplerian matter in the equatorial plane and sub-Keplerian matter flanking it from the top and bottom.
The sub-Keplerian matter suffers shock, puffs in the form of a torus (PSD) and the extra heat evaporates
the Keplerian disc. Various spectral states are easily explained using this hybrid disc model. If the Keplerian
accretion rate increases, the resulting extra soft photons cool down the PSD giving rise to the HSS,
and if the sub-Keplerian accretion rate is higher, it supplies hotter electrons that result in dominant
inverse Comptonized power-law photons as in LHS. All intermediate combinations of the two accretion rates
give rise to other intermediate states. This was later confirmed via numerical simulation \citep{gc13}.
Infact it was shown that the shock in an advective accretion disc shifts towards the BH if viscosity is increased, for a disc of same outer boundary condition \citep{kc13,kc14,kcm14}. As the shock forms closer to the BH, it becomes
stronger, producing faster jets. When radiative moments were computed from the disc and dumped
on to the thermally driven jet,   
\citet{kcm14} showed that as the advective accretion disc moves from LHS to intermediate states, the
mildly relativistic jet becomes much stronger, as was reported in observations \citep{rsfp10}.

Although the advective disc solutions contain features that automatically explain some of the broad observational features (compact corona, compact jet base, spectral state changes), still most of the works
were done in the pseudo-Newtonian regime \citep{pw80,abn96,m03,cm06}.
Full general relativistic treatment were few and far between
\citep{lt80,l85,f87,c96,pa97,ft04,ny08,cc11,ck16}. 
Moreover, although in some of these papers Kerr metric was used,
the issue of jet generation was not addressed \citep[with the exception of][but in Schwarzschild metric]{ck16}. The more favoured model for jet generation in the community
is magnetically driven jets \citep{bp82,c86,fg01}. These relativistic magnetohydrodynamic
solutions were not self-consistently generated from the accretion disc. There are also models that depend
on the extraction of the rotational energy of the BH to power jets \citep{p69,bz77}. Incidentally, the role of BH spin in powering jets has been claimed to be
confirmed \citep{nm12}, as well as been refuted \citep{fgr10,rgf13}. In this paper,
we would like to obtain accretion disc solutions that generate bipolar jets around Kerr BHs and investigate the role
of spin in influencing the jet solution. These results will be valid if the flow is weakly magnetized
or if the magnetic field of 
disc is stochastic in nature.
Since theoretical investigation on jet generation from advective disc around Kerr BH is almost non-existent, therefore, it remains an unattempted question.
It may be noted that \citet{fk07} studied jet generation from advective disc around Kerr BH, but the generation mechanism used was particle
acceleration at the accretion shock. Commendable as the work may be, however, the authors concentrated
their efforts only in estimating the mass-loss and the jet solution was not followed
up to any reasonable distance from the disc: moreover, a fixed $\Gamma$ EoS was also used and hence our approach is quite different. 
One must mention that efforts have been made in estimating mass-loss from advective disc recently with pseudo-Kerr potential \citep{adn15}.
However, since the jet geometry in pseudo-Newtonian potential regime depends weakly on
the BH spin, therefore, one should check the issue of jet generation in full
general relativity, to ascertain the role of spin in influencing jet solution. 

In this paper, we study advective accretion flow around spinning BH.
Although our aim is to study self-consistently launched jets, we present a separate
study of only the
accretion process, before presenting
the simultaneous accretion-ejection solution.
The jets originate from shocked advective discs. The jet geometry is computed from the von  	Zeipel
surfaces \citep[VZS;][]{a71,c85} that we employed previously for jets from accretion discs around Schwarzschild BH \citep{ck16}.
We study the effect of BH spin on the possibility of jet generation, as well as how the jet terminal speed
is affected by the spin. One interesting thing we noticed from earlier accretion-ejection studies is that
the steady, thermally driven
jets obtained in pseudo-Newtonian regime are weak \citep{kc13,kscc13}, but become stronger if
they are powered by radiation \citep{kcm14} or by shock oscillation \citep{lckhr16}.
However, the thermally driven jet is itself stronger in the general relativistic domain compared to the Newtonian regime \citep{ck16}.
Therefore, we would like to investigate whether the jet strength increases with BH spin.
The jet geometry around a spinning BH is decidedly non-spherical. Is it possible to find shocks
in jets mediated by the non-spherical cross-section? Interestingly, for jets around Schwarzschild BH
no such shock in jets were obtained \citep{ck16}. There are some suggestions that
the soft gamma-ray tails in X-ray binaries are due to the presence of shock-accelerated
electrons in the jet, where the shock is situated close to the BH \citep{lrwbpg11}. If
steady shocks in jet are found
to occur close to the BH, then it may support the conclusions of \citet{lrwbpg11}. We would like to explore these questions in details. The novelty of this paper lies in the fact that this is probably the first
effort to investigate all possible jet solutions, including those having multiple sonic points and shocks,
where the jets are computed self-consistently 
from accretion discs.
Moreover, the fluid in accretion-jet system is 
assumed to be fully ionized electron-proton fluid,
described by a relativistic EoS.

In the next section, we present simplifying assumptions and the governing equations. In Section \ref{sec:method},
we outline the solution procedure. In Section \ref{sec:result}, we present the results, and in Section
\ref{sec:conclusn}, we draw concluding remarks.

 \section{Model Equations and Assumptions} \label{sec:eqns}
The estimated temperature of accreting matter is very high: therefore the disc-jet
system is likely to be a fully ionized plasma.
Additionally, we assume the accretion disc-jet system to be axisymmetric. The accretion disc
is assumed to be in hydrostatic equilibrium in the vertical direction and its advection time-scale
is assumed to be shorter than the viscous time-scales. 
Equations of motion of both the accretion disc and the jet
are similar in form, i.e., the
conservation of four-divergence of the respective energy-momentum tensors and the four-mass fluxes.
However, the flow geometries of the jet and accretion disc are quite different,
so the equations of motion of these two entities are presented separately.  

We choose geometrical units $G=\mbh=c=1$, where $G, \mbh$ and $c$ are universal
gravitational constant, mass of the black hole and speed of the light.
Therefore, units of length, speed, angular momentum, energy, time
and mass are $G\mbh/c^2, c,G\mbh/c, \mbh c^2, G\mbh/c^3$ and $\mbh$, respectively.

\subsection{The fluid, its EoS and the background metric}
The energy-momentum tensor of the fully ionized fluid under the present set of assumptions
is given by
\begin{equation}
T^{\mu\nu}=(e+p)u^\mu u^\nu+pg^{\mu\nu},
\label{emt.eq}
\end{equation}
where $e$, $p$ and $u^\mu$ are the local energy density, local gas pressure and
four-velocities, respectively. Here, $g^{\mu\nu}$ are metric components
and indices $\mu, \nu$ are $(0, 1, 2, 3)$.
The governing equations of the relativistic fluids are energy-momentum conservation and in absence
of particle creation, mass-flux conservation given by
\begin{equation}
T_{;\nu}^{\mu\nu}=0, ~~~~~~~~~~~~~~~~~~ (\rho u^\nu)_{;\nu}=0,
\label{consvfld.eq}
\end{equation}
where $\rho$ is the mass density of the flow. The $i$th component of the relativistic Euler equation is obtained by projecting $\tmn_{;\nu}=0$, with the help of projection tensor $h_\mu^i=\delta_\mu^i+u_\mu u^i$,
i. e.
~$h_{\mu}^iT_{;\nu}^{\mu\nu}=0$, or,
\begin{equation}
[(e+p)u^\nu u_{;\nu}^i+(g^{i\nu}+u^iu^\nu)p_{,\nu}]=0,
\label{NS.eq}
\end{equation}
The energy balance equation or entropy equation or the first law of thermodynamics
is $u_\mu T_{;\nu}^{\mu\nu}=0$, or
\begin{equation}
\left[\left(\frac{e+p}{\rho} \right)\rho_{,\mu}-e_{,\mu} \right]=0.
\label{ege.eq}
\end{equation}

A rotating BH is characterized by mass $\mbh$ and spin parameter $a_{\rm s}~(=J/\mbh)$, where $J$ is angular momentum of the BH, since in the present unit system $\mbh=1$: therefore, the analysis presented
in this paper is applicable from stellar mass to supermassive BHs.
The disc-jet system is composed of test fluid that flows in the background metric of
a spinning BH, and is described by Kerr metric and in terms of Boyer-Lindquist coordinates it is given by
\begin{eqnarray}\nonumber
ds^2= g_{\mu\nu}dx^\mu dx^\nu ~~~~~~~~~~~~~~~~~~~~~~~~~~~~~~~~~~~~~~~~~~~~~\\
=g_{tt}dt^2+2g_{t\phi}dtd\phi+g_{rr}dr^2+g_{\theta\theta}d\theta^2+g_{\phi\phi}d\phi^2.
\label{metric.eq}     
\end{eqnarray}
Here, $x^\mu=(x^0, x^1, x^2, x^3)\equiv(t, r, \theta, \phi)$ and $g_{tt}=-(1-2r/\Sigma),~g_{t\phi}=-2a_srsin^2\theta/\Sigma,
~ g_{rr}=\Sigma/\Delta,~ g_{\phi\phi}=(A/\Sigma)sin^2\theta,~ g_{\theta\theta}=\Sigma$ are the
metric components, where, $a_s$ is the spin parameter.
Here, $A=(r^2+a_s^2)^2-\Delta a_s^2sin^2\theta$, $\Sigma=r^2+a_s^2cos^2\theta$ and $\Delta=r^2-2r+a_s^2$.
In the unit system used, the four-velocity components satisfy $u_\mu u^\mu=-1$.
It is clear that for $a_s=0$, the Kerr metric reduces to the Schwarzschild metric. In the other extreme,
$a_s\sim 1$ gives extreme Kerr BHs.
If the flow and the spinning BH are in the same direction, then it is a prograde flow,
i. e. $1\ge a_s>0$ and if they are opposite, then the flow is retrograde or $-1\le a_s<0$.

The equations of motion (equation \ref{consvfld.eq}) are a set of five independent equations
but have six variables.
The fluid follows a relativistic Maxwell-Boltzmann distribution,
which when integrated in the momentum space produces an exact relation between $e$, $p$ and $\rho$
and is called the exact relativistic EoS of the flow \citep{c39,s57,cg68}.
\citet{rcc06,c08} and \citet{cr09} proposed an approximate EoS that is very accurate but much simpler
compared to the exact EoS. The fluid
contains electrons, protons and positrons of different proportions, such that the fluid is
wholly neutral (total charge is zero) and the EoS is given by
\begin{equation}
e=n_{\rm e}m_{\rm e}c^2f=\frac{\rho}{\tau} f,
\label{reos.eq}
\end{equation}
where, $\rho(=\rho_{\rm e}\tau=n_{\rm e}m_{\rm e}\tau)$ is the total mass density (see details in \citealt{c08,cr09}), $\tau=[2-\xi(1-1/\chi)]$ and 
\begin{equation}
f=(2-\xi)\left[1+\Theta\left(\frac{9\Theta+3}{3\Theta+2}\right)\right]+\xi\left[\frac{1}{\chi}+
\Theta\left(\frac{9\Theta+3/\chi}{3\Theta+2/\chi}\right)\right].
\label{f.eq}
\end{equation}
Here, $\xi=n_{\rm p}/n_{\rm e}$ is the composition parameter, 
$\Theta=(kT)/(m_{\rm e}c^2)$ is the dimensionless temperature of the fluid and 
$\chi=m_{\rm e}/m_{\rm p}$ is the mass ratio, i. e. ratio between the electron mass and the proton mass.
Here, $n_{\rm p}$ and $n_{\rm e}$ are the proton 
number density and the electron number density, respectively.
The equivalence of the different forms of exact relativistic EoS \citep{c39,s57,cg68}
and its comparison with the Chattopadhyay$-$Ryu approximate EoS
show a very close agreement \citet{vkmc15}.
In the relativistic case, polytropic index ($N$), the adiabatic index 
($\Gamma$) and sound speed ($a$) are defined as,
\begin{equation}
N=\frac{1}{2}\frac{df}{d\Theta};~~~\Gamma=1+\frac{1}{N} ~~~\mbox{and}~~~a^2=\frac{\Gamma p}{e+p}=\frac{2\Gamma\Theta}{f+2\Theta}.
\label{nga.eq}
\end{equation}
In this paper, we have investigated only the electron$-$proton case i.e., $\ep$ or $\xi=1.0$.
In the following, we present the equations of motion of the accretion disc and the jet separately.

\subsection{Accretion Equations of motion}\label{subsec:aeqns}
The accretion disc occupies the space around the equatorial plane of the BH. The equations of motion for the 
accretion disc are written on the equatorial plane (i.e., $\theta \rightarrow \pi/2$).
The motion along the transverse direction is negligible i.e., $u^{\theta}=0$. The flow variables
are considered on the equatorial plane  and assumed to remain constant along the transverse direction (i. e., $\partial/\partial \theta =0$) while axisymmetry renders $\partial/\partial \phi=0$.

The radial component of momentum balance equation (\ref{NS.eq}), is
\begin{eqnarray}\nonumber
u^r\frac{du^r}{dr}+\frac{1}{2}g^{rr}g_{tt,r}g^{tt}+\frac{1}{2}u^ru^r\left[g^{tt}g_{tt,r}+
g^{rr}g_{rr,r}\right]-\\ \nonumber
\frac{1}{2}g^{rr}g_{tt,r}g^{t\phi}u^tu_\phi+\frac{1}{2}g^{rr}u^\phi u^\phi\left[g_{\phi\phi}g^{tt}g_{tt,r}-g_{\phi\phi ,r}\right]+\\ 
\frac{1}{2}g^{rr}u^tu^\phi\left[g_{t\phi}g^{tt}g_{tt,r}-2g_{t\phi ,r}\right]
+\frac{(g^{rr}+u^ru^r)}{e+p}\frac{dp}{dr}=0.
 \label{rNS.eq}
 \end{eqnarray}
The integrated form of the azimuthal component ($h_{\mu}^{\phi}T_{;\nu}^{\mu\nu}=0$) of energy-momentum conservation of
equation (\ref{consvfld.eq}) is
\begin{equation}
L=hu_{\phi}=\mbox{constant}=hl;~~ h=\frac{e+p}{\rho}=\frac{f+2\Theta}{\tau},
 \label{phiNS.eq}
\end{equation}
where $h$ is the specific enthalpy, $L$ is the bulk angular momentum of the flow per unit mass, while the definition of
specific angular momentum is $\lambda=-u_{\phi}/u_t$. The covariant $\phi$ component of the four-velocity
is also represented as $l=u_\phi$. One may also obtain equation (\ref{phiNS.eq}) with the aid of an
azimuthal Killing vector $\zeta^\mu_\phi=(0,0,0,1)$.
The $\theta$ component of momentum balance equation (\ref{NS.eq}) can be integrated assuming hydrostatic equilibrium around the equatorial
plane (\ie $~\theta \approx \pi/2$) to obtain local disc half-height expression
\citep{rh95,pa97},
\begin{equation}
H=rH_\theta={\left(\frac{p r^3}{\rho{\cal F}}\right)}^{1/2},
 \label{hhe.eq}
\end{equation}
where ${\cal F}=\gamma_{\phi}^2[{(r^2+a_s^2)^2+2\Delta a_s^2}][{(r^2+a_s^2)^2-2\Delta a_s^2}]^{-1}$.
Integrating conservation of mass flux in equation (\ref{consvfld.eq}), we obtain
the expression of accretion rate,
\begin{equation}
 \dot{M}=\int r^2 \rho u^r d\theta d\phi = 4\pi r^2 H_\theta \rho u^r=4\pi\rho H u^r r.
 \label{mc.eq}
\end{equation}
If one integrates equation (\ref{ege.eq}), we obtain the adiabatic EoS for relativistic
multispecies adiabatic flow with the help of EoS (equation \ref{reos.eq}) as \citep{ck13,ck16,kscc13}
\begin{equation}
\rho={\cal K}e^{k_3}\Theta^{3/2}(3\Theta+2)^{k_1}(3\Theta+2/\chi)^{k_2},
 \label{peos.eq}
\end{equation}
where ${\cal K}$ is the entropy constant and $k_1=3(2-\xi)/4, k_2=3\xi/4, k_3=(f-\tau)/(2\Theta)$. This relativistic equation is 
analogous to non-relativistic polytropic EoS, $p={\cal K}\rho^{\Gamma}$. Using equation (\ref{peos.eq})
in equation (\ref{mc.eq}), we get 
entropy-accretion rate,
\begin{equation}
\dot{\cal M}=\frac{\dot{M}}{4\pi{\cal K}}=u^rHre^{k_3}\Theta^{3/2}(3\Theta+2)^{k_1}(3\Theta+2/\chi)^{k_2},
 \label{enaccn.eq}
\end{equation}
$\dot{\cal M}$ is a constant for adiabatic flow.

Since $\partial g_{\mu \nu}/\partial t=0$, so one may obtain a constant of motion with the help of
a time$-$like Killing vector $\zeta^\mu_t=(1, 0, 0, 0)$ in the equation of motion we have
$$
(\zeta^{\mu}_tT^{\nu}_{\mu})_{;\nu}=\frac{1}{\sqrt{-g}}\left({\sqrt{-g}}T^r_t\right)_{,r}=0.
$$ 
This equation when integrated give us the negative of the energy flux,
\be
4\pi H_\theta r^2\left[(e+p)u^ru_t \right]={-\dot E}.
\label{energflux.eq}
\ee
 The relativistic Bernoulli equation is obtained by combining equations (\ref{mc.eq}) and
 (\ref{energflux.eq}),
\begin{equation}
 {\cal E}=\frac{\dot E}{\dot M}=-hu_t=\mbox{constant,}
 \label{Ber.eq}
\end{equation}
Here, $u_t=\sqrt{1/(1-\omega_\phi\lambda)}\alpha\gamma$, where $\gamma=\gamma_v\gamma_\phi$ is the
total bulk Lorentz factor, 
$\gamma_\phi=1/\sqrt{1-\Omega\lambda}$ is the bulk azimuthal Lorentz factor and $\gamma_v=1/\sqrt{1-v^2}$ is the bulk radial Lorentz factor. 
Moreover, $\alpha^2=(r^2\Delta)/A$ and $\omega_\phi=-g_{t\phi}/g_{\phi\phi}$. The angular velocity is $\Omega=u^\phi/u^t$.

We simplify equation (\ref{rNS.eq}) with the help of equations (\ref{ege.eq}), (\ref{hhe.eq}, (\ref{mc.eq}  and (\ref{reos.eq}) and get 
derivative of bulk velocity,
\begin{equation}
 \frac{dv}{dr}=\frac{\cal N}{\cal D},
 \label{dvdr.eq}
\end{equation}
where
\begin{eqnarray*}
{\cal N} & = &2Na^2\{(2N+1)\Delta\}^{-1}\left(1-a_s^2/r\right)-A(r^4\Delta)^{-1} \\
&- &
A\gamma_\phi^2(r^2\Delta)^{-1}(1-w_\phi\lambda)\frac{\Omega^2}{\alpha^2}\left[\frac{A}{r^4}-
\alpha^2\left(r-\frac{a_s^2}{r^2}\right)\right] \\ 
& + & \gamma_\phi^2(1-w_\phi\lambda)\frac{\Omega}{\alpha^2}\frac{2a_s}{r^2\Delta}
\left[\frac{A}{r^3}+\Delta\right]+\frac{2a_s\lambda}{r^3\Delta}\gamma_\phi^2 \\
& + &\frac{Na^2}{(2N+1)}\left[\frac{5}{r}-
\frac{1}{{\cal F}}\frac{d{\cal F}}{dr}\right]
\end{eqnarray*}
and 
$$
{\cal D}=\frac{\gamma_{v}^2}{v}\left[v^2-\frac{2Na^2}{(2N+1)}\right].
$$

And from equations (\ref{ege.eq}), we get derivative of dimensionless temperature,
\begin{equation}
\frac{d\Theta}{dr}=-\frac{2\Theta}{(2N+1)}\left[\frac{1}{\Delta}
\left(1-\frac{a_s^2}{r}\right)+\frac{\gamma_v^2}{v}\frac{dv}{dr}+
\frac{5}{2r}-\frac{1}{2{\cal F}}\frac{d{\cal F}}{dr}\right],
 \label{dthdr.eq}
\end{equation}
where
$\frac{1}{{\cal F}}\frac{d{\cal F}}{dr}=\frac{\lambda\gamma_\phi^2}{(A-2a_sr\lambda)}\left[2(\lambda-a_s)-
\Omega(A^{'}-2A/r+2a_s\lambda)\right]+\frac{A^{'}+3\Delta^{'}a_s^2}{A+3\Delta a_s^2}-\frac{A^{'}}{A}$, $A^{'}=4r^3+2ra_s^2+2a_s^2$ 
and $\Delta^{'}=2(r-1)$.

\subsubsection{Accretion critical points}\label{subsubsec:accncp}
BH accretion is necessarily transonic because the inner boundary condition is always supersonic.
At $r\rightarrow r_{\rm c}$ the critical point, $dv/dr\rightarrow 0/0$, which gives the critical point
conditions, 
\begin{equation}
 {\cal N}=0~~\Longrightarrow~~
v_{\rm c}^2=\frac{\frac{A_{\cm}}{r_{\cm}^4\Delta_{\cm}}+{\rm N}_{\rm u}}
{\frac{1}{\Delta_{\rm c}}(1-\frac{a_s^2}{r_{\rm c}})+\frac{5}{2r_{\rm c}}-\frac{1}{2{\cal F}_{\cm}}\frac{d{\cal F}_{\cm}}{dr}},
 \label{cn.eq}
\end{equation}
\begin{eqnarray}\nonumber
{\rm N}_{\rm u}=
 \frac{A_{\cm}\gamma_{\phi\cm}^2}{r_{\cm}^2\Delta_{\cm}}(1-w_{\phi\cm}\lambda)\frac{\Omega_{\cm}^2}{\alpha_{\cm}^2}
 \left[\frac{A_{\cm}}{r_{\cm}^4}-\alpha_{\cm}^2(r_{\cm}-\frac{a_s^2}{r_{\cm}^2})\right]-\\ \nonumber
 \gamma_{\phi\cm}^2(1-w_{\phi\cm}\lambda)\frac{\Omega_{\cm}}{\alpha_{\cm}^2}\frac{2a_s}{r_{\cm}^2\Delta_{\cm}}
\left[\frac{A_{\cm}}{r_{\cm}^3}+\Delta_{\cm}\right]-\frac{2a_s\lambda}{r_{\cm}^3\Delta_{\cm}}\gamma_{\phi\cm}^2
\end{eqnarray}
and
\begin{equation}
 {\cal D}=0~~~\Longrightarrow~~~v_{\rm c}^2=\frac{2N_{\rm c}a_{\rm c}^2}{2N_{\rm c}+1},
 \label{cd.eq}
\end{equation}
where subscript `c' denotes a flow variable at critical point. The gradient of velocity
at $r_{\cm}$ is obtained by L'H\^opital's rule.
Relativistic Bernoulli parameter (equation \ref{Ber.eq}) at critical point is written as
\begin{equation}
 {\cal E}=-hu_t=-h_cu_{t\cm}=h_{\cm}\alpha_{\cm}\frac{\sqrt{1/(1-w_{\phi\cm}\lambda)}}{\sqrt{(1-v_{\cm}^2)(1-\Omega_{\cm}\lambda)}},
 \label{ecri.eq}
\end{equation}
Solving equations (\ref{cn.eq})$-$(\ref{ecri.eq}) together, we find that for a given set of parameters (${\cal E},~\lambda,~\xi$ and $a_s$), flow may have single or two to three critical points.
In case there are three critical points, then the inner ($\rci$) and outer ($\rco$) critical points are of saddle type (or $X-$type) and middle ($\rcm$) is of the centre type
($O-$type). Therefore, the accretion solutions, which come from infinity, on to the BH horizon, can pass only
through $\rci$ or $\rco$ or both critical points when shock transition occurs \citep{f87}.
\subsection{Equations of motion for jets}\label{subsec:jeqns} 
Although the general form of jet equations of motion is the same as that of the accretion disc
(equation \ref{consvfld.eq}), the flow geometry is entirely different. The accretion
disc occupies the space around the equatorial plane ($\theta=\pi/2$), but the jet flows about the axis of symmetry. Therefore, none of the three-velocity
components of the jet are negligible. If we define the coordinate velocities and respective 
momentum per inertial mass as \citep{c85}
\begin{equation}
\vartheta_{\rj}^{i}=\frac{u_{\rj}^i}{u^t_{\rj}} ~~~\mbox{and}~~~ 
\vartheta_{i\rj}=-\frac{u_{i{\rj}}}{u_{t{\rj}}},
\label{jtol.eq}
\end{equation}
where indices $i=(r, \theta, \phi)$ and subscript `{\rj}' represents the quantities for the jet flow.
The azimuthal three$-$velocity of the jet $v_\phi^2=\vartheta_{\phi \rj}\vartheta_{\rj}^\phi$
will be orthogonal to the stream line, and the advection three$-$velocity $v_p^2=\vartheta_{r \rj}\vartheta^r_{\rj}+\vartheta_{\theta \rj}\vartheta_{\rj}^\theta$ will be along the streamline.
Here, the jet angular velocity and specific angular momentum are defined as 
$\Omega_{\rj}=\vartheta_{\rj}^{\phi}$ and
$\lambda_{\rj} = \vartheta_{\phi \rj}$, respectively.
Above the equatorial plane the constant angular momentum surface is the VZS,
which are also the surfaces of constant entropy. VZS is characterized by von Zeipel parameter
given by \cite{kja78} and \cite{c85},
\begin{equation}
Z_{\phi}^2=\frac{\vartheta_{\phi\rj}}{\vartheta_{\rj}^{\phi}}=
\left[\frac{A_{\rj}-2a_sr_{\rj}\lambda_{\rj}}{\Delta_{\rj}-a_s^2sin^2\theta_{\rj}
	+2a_sr_{\rj}sin^2\theta_{\rj}/\lambda_{\rj}}\right]sin^2\theta_{\rj}~ ,
\label{vZp.eq}
\end{equation}
where $A_{\rj},~a_s,$ and $\Delta_{\rj}$ are the properties of the metric (see equation \ref{metric.eq}),
applied to the jet.
\citet{c85} showed that in order to obtain the streamline, the following relation should hold,
 \begin{equation}
  \vartheta_{\phi\rj}=c_{\phi} Z_{\phi}^{n_{\cm}}, 
  \label{lvzp.eq}
 \end{equation}
where $c_{\phi}$ and $n_{\cm}$ are some constant parameters. \citet{ck16} connected these jet streamlines
with the accretion disc, where,, $n_{\rm c}$ and $c_{\phi}$
were determined from the accretion disc properties.
If $\rsh$ is the location of shock and $\rci$ is the inner sonic
point of the accretion disc, then the radius of jet base on the equatorial plane is $x_\rb=(\rci+\rsh)/2$,
and the polar angle of the jet base on the disc surface is $\theta_{\rb}={\rm tan}^{-1}\left(x_\rb/H_{\rb}\right)$, where $H_\rb$ is the
disc height at $x_\rb$. Then the radius of the jet base is $r_\rb= x_\rb {\rm cosec}\theta_\rb$.
The cross-section area of jet, orthogonal to the jet streamline at $r_{\rj}$ is given by
\begin{equation}
{\cal A}_{\rj}={\cal A}_{\rb} \left(\frac{r_{\rj}}{r_{\rb}}\right)^2 {\rm sin}\theta_{\rj},
\label{jtar.eq}
\end{equation}
where ${\cal A}_{\rb}={\cal A}_{\rb}^\prime {\rm sin}\theta_{\rb}$ and ${\cal A}_{\rb}^\prime=2\pi(r_{\rb 0}^2-r_{\rb{\ri}}^2)$. Here,
$r_{\rb{\ri}}=\rci/{\rm sin}\theta_\rb, r_{\rb 0}=
\rsh/{\rm sin}\theta_\rb$. 
And, $\theta_{\rj}$ is defined in Appendix \ref{app:gpp}.
If the equations of motion of the jet are integrated along the streamline (equation \ref{lvzp.eq}),
along with the EoS (equation \ref{reos.eq}) and equations (\ref{vZp.eq}) and (\ref{lvzp.eq}), a constant
of motion that is similar to the Bernoulli parameter is obtained,
\begin{equation}
 {\cal R}_{\rj}=-h_{\rj}u_{t\rj}[1-c_\phi^2Z_\phi^{(2n_{\cm}-2)}]^\beta={\cal E}_{\rj}[1-c_\phi^2Z_\phi^{(2n_{\cm}-2)}]^\beta={\cal B}_{\rj}\gamma_{\phi \rj}^{(1-2\beta)}.
 \label{Berj.eq}
\end{equation}
Here, $u_{t\rj}=-\sqrt{1/(1-\omega_{\phi\rj}\lambda_{\rj})}\alpha_{\rj}\gamma_{\rj}$, 
$\alpha_{\rj}=(\Delta_{\rj}\Sigma_{\rj})/(A_{\rj}sin^2\theta_{\rj})$, $\gamma_{\rj}=\gamma_{v\rj}\gamma_{\phi\rj}$, 
$\gamma_{v\rj}=1/\sqrt{1-v_{\rj}}$, $\gamma_{\phi\rj}=1/\sqrt{1-c_\phi^2Z_\phi^{(2n_{\cm}-2)}}$,
$\beta=n_{\cm}/(2n_{\cm}-2)$, ${\cal E}_{\rj}$ is the jet Bernoulli parameter and 
the jet velocity measured in corotating frame is $v_{\rj}=\gamma_{\phi\rj}v_p$ \citep[see][for definition]{l85}.
In equation (\ref{Berj.eq}), ${\cal B}_{\rj}$ is the effective Bernoulli parameter, obtained after extracting all the
$\gamma_{\phi \rj}$ terms from the expression of ${\cal R}_{\rj}$.

Integral form of continuity equation gives mass outflow expression along the streamline that
can be written as
\begin{equation}
 \dot{M}_o=\rho_{\rj}u_{\rj}^p {\cal A}_{\rj},
 \label{mof.eq}
\end{equation}
where $u_{\rj}^p=\sqrt{g^{pp}}\gamma_{v\rj}v_{\rj}$ is the jet four-velocity parallel to the jet stream line, $\rho_{\rj}$ is the mass density
and ${\cal A}_{\rj}$ is the
area of the jet cross-section orthogonal to the streamline, respectively. The $g^{pp}=1/h_p^2$ is defined in Appendix \ref{app:gpp}. 
Similar to the entropy-accretion rate equation, we can also define the entropy-outflow rate for the jet flow, 
\begin{equation}
 \dot{\cal M}_{\rj}=u_{\rj}^p{\cal A}_{\rj}e^{k_3}\Theta_{\rj}^{3/2}(3\Theta_{\rj}+2)^{k_1}(3\Theta_{\rj}+2/\chi)^{k_2},
 \label{enjt.eq}
\end{equation}
This entropy-outflow rate quantity is also constant along the jet streamline except at the shock.

The differential form of equations (\ref{Berj.eq}) and (\ref{mof.eq}) with the help of
equation (\ref{peos.eq}) can be 
expressed as derivatives of jet velocity and dimensionless temperature,
\begin{equation}
\frac{dv_{\rj}}{dr_{\rj}}=\frac{{\cal N}_{\rj}}{{\cal D}_{\rj}}=
\frac{{a_{\rj}^2}{({\cal A}_{\rj})^{-1}}{d{\cal A}_{\rj}}/{dr_{\rj}}-
	{a_{\rj}^2}{h_p}^{-1}{dh_p}/{dr_{\rj}}-X_{\rm g}}
{v_{\rj}\gamma_{v\rj}^2[1-{a_{\rj}^2}/{v_{\rj}^2}]} ,
 \label{dvdrj.eq}
\end{equation}
and 
\begin{equation}
\frac{d\Theta_{\rj}}{dr_j}=
-\frac{\Theta_{\rj}}{N_{\rj}}\left[\frac{\gamma_{v\rj}^2}{v_{\rj}}\frac{dv_{\rj}}{dr_{\rj}}+
\frac{1}{{\cal A}_{\rj}}\frac{d{\cal A}_{\rj}}{dr_{\rj}}-\frac{1}{h_p}\frac{dh_p}{dr_{\rj}} \right],
 \label{dthj.eq}
\end{equation}
where
$$X_{\rm g}=\frac{0.5}{(g_{tt}+w_{\phi}g_{t\phi})}
\left[\frac{dg_{tt}}{dr_{\rj}}+\frac{d(w_{\phi}g_{t\phi})}{dr_{\rj}}\right]+
\frac{\lambda_{\rj}}{2(1-w_{\phi}\lambda_{\rj})}\frac{dw_{\phi}}{dr_{\rj}}.
$$
If we compare the jet equation in the present paper with radial outflow equations
around Schwarzschild metric, then by making transformations such as
$u^p_{\rj}\rightarrow u^r$,
 ${\cal A}_{\rj}^{-1}d{\cal A}_{\rj}/dr_{\rj}\rightarrow 2/r$,$h_p\rightarrow h_r$,
 we will be able to transform 
 equation (\ref{dvdrj.eq}) to equation 7a of \citet{cr09}. This allows us to identify
the first term of ${\cal N}_{\rj}$ as the coupling between cross-section and the thermal term,
the second one being the coupling between gravity and the thermal term and the third term or
$X_{\rm g}$ as purely the gravity term.

\subsubsection{Jet critical points}\label{subsubsec:jtcp}
Similar to accretion critical points, we can also find jet critical points by using ${\cal N}_{\rj}={\cal D}_{\rj}=0$, which gives 
two critical point conditions and they are
\begin{equation}
v_{\rj c}=a_{\rj\cm}~~~\mbox{and}~~~ v_{\rj\cm}=
\sqrt{\frac{X_{\rm g}}{ {{\cal A}^{-1}_{\rj\cm}} {d{\cal A}_{\rj\cm}}/{dr_{\rj\cm}}-{h^{-1}_p}{dh_p}/{dr_{\rj\cm}}}} 
 \label{jtcp.eq}
\end{equation}
At the jet sonic point, i. e. $r_{\rj\cm}$, the gradient of velocity is determined by L'H\^opital's rule.

\subsection{Shock waves in accretion and bipolar jets}\label{subsec:shockeqn}
It has been shown previously that accretion disc shock may launch bipolar jets
\citep{mlc94,msc96,mrc96,dcnm14,lckhr16}. 
The jump condition across an accretion shock is obtained by conserving fluxes across
the shock front, \eg, mass flux, momentum flux and energy flux \citep{t48}.
In the presence of mass-loss at the shock, mathematically they are
\begin{equation}
 \dot{M}_+=\dot{M}_--\dot{M}_{o}=\dot{M}_-(1-R_{\dot{m}}),
 \label{mf.eq}
\end{equation}
where $R_{\dot{m}}=\dot{M}_o/\dot{M}_{-}$ is the relative mass outflow rate,
\begin{equation}
 H_+\rho_+h_+\gamma_{v+}^2v_+^2+p_+H_+=H_-\rho_-h_-\gamma_{v-}^2v_-^2+p_-H_-,
 \label{momf.eq}
\end{equation}
and
\begin{equation}
 {\cal E}_+={\cal E}_-.
 \label{ef.eq}
\end{equation}
Here, subscripts `$-$' and `$+$' are representing flow quantities before and after shock transition, respectively. 
Now, we have solved three shock conditions (\ref{mf.eq})$-$(\ref{ef.eq}) simultaneously, where angular momentum is continuous across 
the shock and we get the relation between pre-shock and post-shock flow variables as
\begin{equation}
 h_{-}^{\prime}u_{-}^2-{\rm k}_1u_{-}+2\Theta_{-}=0~~\mbox{and}~~{\rm k}_2-h_{-}^{\prime}\gamma_{v-}=0,
\label{accjump.eq}
\end{equation}
where ${\rm k}_1=(1-R_{\dot{m}})(h_+^\prime u_+^2+2\Theta_+)/u_+, 
~{\rm k}_2=h_+^\prime{\gamma_v}_+,~h_{\pm}^\prime=(f_{\pm}+2\Theta_{\pm})~\mbox{and}~u=v\gamma_v$.
The explicit expression of relative mass outflow rate was defined
with the help of equations (\ref{mc.eq}), (\ref{jtar.eq}) and (\ref{mof.eq}) 
in terms of post-shock flow variables and jet base quantities and is written as \citep{ck16}
\be
\begin{split}
R_{\dot{m}}=\frac{\dot{M}_o}{\dot{M}_{-}}=\frac{1}{\left[{(4\pi H_+r_+\rho_+u_+^r)}/{({\cal A}_{{\rj}\rb}\rho_{{\rj}\rb}u_{{\rj}\rb}^p)}+1 \right]} \\ \nonumber
= \frac{1}{\left[\varSigma{(R_{\rm A}R\varXi)}^{-1}+1 \right]},
\end{split}
\label{rmd2.eq}
\ee
where $\rho_{{\rj}\rb}=\rho_\rb {\rm exp}(-7x_\rb/(3h_\rb))/h_\rb^2, u_{{\rj}\rb}^p=\sqrt{g^{pp}}\gamma_{v\rb}v_{{\rj}\rb}$ and 
${\cal A}_{{\rj}\rb}={\cal A}_\rb {\rm sin}\theta_\rb$ are the jet base density,
the four-velocity at jet base and the
cross-section area of jet base, respectively.
Moreover, $R_{\rm A}={\cal A}_{{\rj}\rb}/(4\pi H_+r_+)$, $R=(u_-^r)/(u_+^r)$ the compression ratio, $\varSigma=\rho_+/\rho_-$
the density jump across the accretion shock 
and $\varXi=(\rho_{{\rj}\rb}u_{{\rj}\rb}^p)/(\rho_-u_-^r)$ the ratio of the relativistic mass flux of the jet base
to the pre-shock accretion flow, respectively. Therefore, $\rmdot$ will increase with
increasing $R,~R_{\rm A}$ and $\varXi$, but will decrease with increasing $\varSigma$. 

\subsubsection{Shock in jets}
Parallel to conservation of fluxes of conserved quantities in accretion disc, we can also conserve the fluxes of jet flows along the streamline to obtain shock conditions,
\begin{equation}
[\dot{M}_{o}]=0,~~~[\rho_{\rj}h_{\rj}\gamma_{v\rj}^2v_{\rj}^2+p_{\rj}]=0~~~\mbox{and}~~~[{\cal R}_{\rj}]=0
 \label{sokjt.eq}
\end{equation}

\section{Solution Procedure}\label{sec:method}
In this paper, we obtain accretion-jet solutions self$-$consistently, similar to our previous effort
in Schwarzschild metric \citep{ck16},
but now for Kerr metric. Recently, we showed that smooth accretion solutions do not automatically drive
bipolar jets, while shocked accretion may drive jet from the PSD \citep{lckhr16}. Therefore, we look for
shocked accretion solution. Once we obtain shocked accretion solution, then we search for the transonic
jet solutions. Moreover, we also check for
shocks in jets. In the following, for a given $a_s$ and $\xi$, we list the exact methodology to obtain the accretion-jet solution. \\
\begin{enumerate}
	\item[(i)] We supply ${\cal E}$ and $\lambda$ in equation (\ref{ecri.eq}) and with the help of critical point conditions (equations \ref{cn.eq} and \ref{cd.eq}) we obtain the critical points of the flow. In a significant part of the ${\cal E}$---$\lambda$ parameter space, multiple critical points (MCP) are possible. 
	\item[(ii)] We determine the gradient of $v$ at the critical point by employing L'H\^opital's rule. Then starting from the critical points we integrate equations (\ref{dvdr.eq}) and (\ref{dthdr.eq}) to obtain the solutions.
	\item[(iii)] For those ${\cal E}$ and $\lambda$ that admit MCP, the entropy at each such
	point is checked. If the entropy at $\rco$ $\eno$ is less than $\eni$ at $\rci$, then we check for shock jump (equation \ref{accjump.eq}) considering $R_{\dot m}=0$. If there is a shock, supersonic flow
	through $\rco$ jumps to the subsonic branch, which ultimately becomes transonic at $\rci$ and falls into the
	BH as supersonic flow.
	\item[(iv)] Once the shock in accretion $\rsh$ is found, we assume the base of the
	jet streamline to be at $r_{\rm b}=(\rci+\rsh)/2$. At $r_{\rm b}$, we compute ${\cal R}_{\rj}$ of the jet
	from the disc values of ${\cal E}$ and $\gamma_{\phi}$ and a guess value of $n_c$.
We then evaluate the von Zeipel parameter (equation \ref{vZp.eq}), as we did
for the Schwarzschild case \citep{ck16}, but now for Kerr metric. And then using these
values we look for transonic jet solution.
We iterate $c_{\phi}$ and $n_{\cm}$ for the same values of $Z_\phi$ and $r_{\rm b}$
and by estimating the value of ${\cal R}_{\rj}$ from the disc. This goes on until
the entropy of the transonic jet is in between the PSD and pre-shock disc or ${\cal {\dot M}}_{\rci} > {\cal {\dot M}}_{\rj} > {\cal {\dot M}}_{\rco}$.
	\item[(v)] Once the transonic solution is found, then from Section \ref{subsec:shockeqn}, we compute the
	$R_{\dot m}$, which is fed to equation (\ref{accjump.eq}) to recalculate the $\rsh$. Once the new $\rsh$ is 
	obtained, steps (iii)$-$(v) are repeated but with the new value of $\rmdot$, till $\rsh$ converges to a value. The resulting jet
	is the self-consistent jet solution driven by disc properties.
	\item[(vi)] While the jet solution is being obtained, shock transition in jets is also studied.
\end{enumerate} 
\section{Results}\label{sec:result}
We are connecting two flows, namely accretion and jet, and since both the flows are
quite complicated, we would first present all possible advective accretion solutions, then all possible jet solutions
and then self-consistent accretion-jet solutions.
\subsection{Accretion}\label{subsec:accret}
\begin{figure}
\centering
\includegraphics[width=8.0cm]{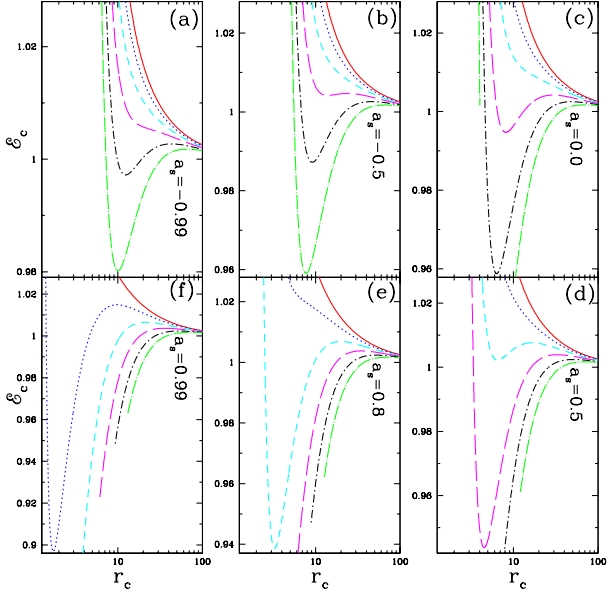}
\caption{Accretion disc critical point properties: ${\cal E}_{\cm}$ as a function of critical point $r_{\cm}$ for
	different Kerr spin parameters, 
	$a_s=-0.99$ (a), $a_s=-0.5$ (b), $a_s=0.0$ (c), $a_s=0.5$ (d), $a_s=0.8$ (e) and $a_s=0.99$ (f).
	Each of the curves from top to bottom is for $\lambda=1.5 - 4.0$ with interval $d\lambda=0.5$. 
	All curves are for the same composition parameter, $\xi=1.0$ or $\ep$ flow.}
\label{fig:fig1}
\end{figure}
In this section, we present only accretion solution, i. e. we follow steps (i)$-$(iii) of
Section \ref{sec:method}.
BH accretion disc is necessarily transonic, i. e. the solutions falling on to a BH passes
through atleast one sonic point. Therefore, a lot of insight can be gained by studying the
critical point conditions.
For a given value of ${\cal E}$ and $\lambda$,
the critical points can be uniquely determined. Let us denote ${\cal E}_{\cm}={\cal E}|_{r_{\cm}}$ and ${\dot {\cal M}}_{\cm}={\dot {\cal M}}|_{r_{\cm}}$. Therefore, using the critical point conditions
(equations \ref{cn.eq} and \ref{cd.eq}) in equation(\ref{Ber.eq}), we obtain energy as a function
of $r_c$ for given values of $\lambda$, $a_s$ and $\xi$ (equation \ref{ecri.eq}).
We plot ${\cal E}_{\cm}$ with $r_{\cm}$ for $a_s=-0.99$ (Fig.\ref{fig:fig1}a), $a_s=-0.5$ 
(Fig.\ref{fig:fig1}b), $a_s=0.0$ (Fig.\ref{fig:fig1}c), $a_s=0.5$ (Fig.\ref{fig:fig1}d),
$a_s=0.8$ (Fig.\ref{fig:fig1}e) and $a_s=0.99$ (Fig.\ref{fig:fig1}f).
Each of the curves plotted from top to bottom is for $\lambda=1.5 - 4.0$ with interval $d\lambda=0.5$
and with same composition parameter, $\xi=1.0$. Critical points are formed because gravity
increases the infall velocity, but because accretion is an example of convergent flow the temperature of the flow also increases during accretion (compressional heating).
As temperature increases, sound speed also increases, but in accretion the rise of sound speed
is less than the bulk velocity.
Therefore, the flow that is subsonic at large distances
becomes supersonic within a certain point (critical point). However, if the infalling matter is rotating,
then the centrifugal term competes with the gravitation term in a way that MCP may form.
As $a_s$ increases, frame dragging enhances the centrifugal term significantly, therefore,
even matter with low angular momentum exhibits MCP, for example, the flow with $\lambda=2$
[dotted in Figs. \ref{fig:fig1}a-f] exhibits MCP for $a_s=0.99$ but not for $a_s\leq 0.8$. On the other hand, if the flow is retrograde, then MCP occur only for higher angular momentum flow.
Fig. (\ref{fig:fig1}c) is similar to fig. 3a of \citet{cc11}. This phenomenon is better exhibited
with ${\cal E}_{\cm}-{\dot {\cal M}}_{\cm}$ `kite-tail' plot. 
\begin{figure}
\centering
\includegraphics[width=7.0cm]{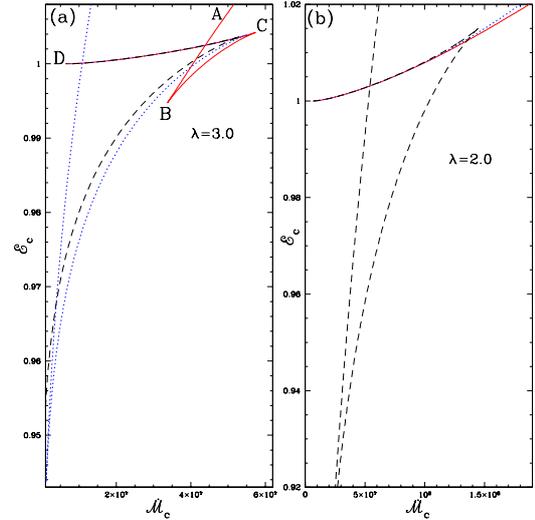}
\caption{Accretion disc critical point properties: ${\cal E}_{\cm}$ versus ${\dot {\cal M}}_{\cm}$; each curve
	plotted for Kerr parameters, 
	$a_s=0$ (solid, red), $a_s=0.5$ (dotted, blue), $a_s=0.99$ (dashed, black). (a) Angular momentum $\lambda=3$ (b) $\lambda=2$.
	All curves are for the same composition parameter, $\xi=1.0$ or $\ep$ flow.}
\label{fig:fig2}
\end{figure}

For a given set of ${\cal E}$ and $\lambda$, there can be one to three critical points.
In general, when MCP occur for a given set of ${\cal E}~\&~\lambda$, then each of the critical points possesses different entropy from the other.
In ${\cal E}_{\cm}-{\dot {\cal M}}_{\cm}$ (i.e. evaluated at critical points) for a given $\lambda$,
the curve takes the shape of a kite-tail. In Fig. \ref{fig:fig2}(a), we plot ${\cal E}_{\cm}$ versus
${\dot {\cal M}}_{\cm}$ curves for $a_s=0$ (solid, red), $a_s=0.5$ (dotted, blue) and $a_s=0.99$ (dashed,
black) all the curves for the same angular momentum $\lambda=3.0$. Branch AB (marked on the
curve related to $a_s=0$)
corresponds to $\rci$, BC corresponds to $\rcm$ and CD to $\rco$. This implies that, for a given $\lambda$,
there is a range of ${\cal E}$ and ${\dot {\cal M}}$ for which MCP may form. As we opt for higher spin of the BH
(dotted, dashed),
the MCP region shifts to the lower entropy region. And for highly spinning BH, for example
$a_s=0.99$, the inner critical point ($\rci$) does not form.  In contrast, for flows with lower $\lambda~(=2)$, MCP form for highly spinning BH like
$a_s=0.99$. But for lower $a_s$, accretion flows with low $\lambda$ form only $\rco$-type sonic points
(Fig. \ref{fig:fig2}b).
Interestingly, Figs. (\ref{fig:fig2}a) and (b) show that, for a given
${\cal E}_c$ and ${\dot {\cal M}_c}$, the outer sonic points or $\rco$ are formed at roughly
the same location in accretion discs around BHs of
different $a_s$, but the locations of $\rcm$
and $\rci$ differ widely.

\begin{figure}
\centering
\includegraphics[width=8.0cm]{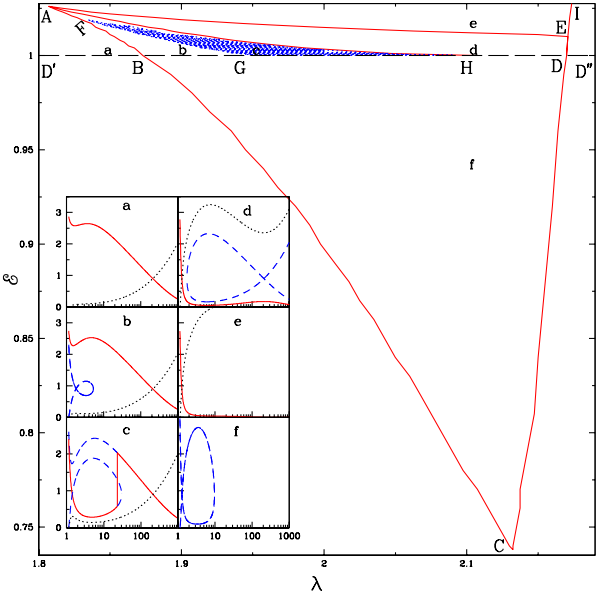}
\caption{Accretion disc: ${\cal E}-\lambda$ parameter space, showing single sonic point, MCP and shock region,
with typical solutions from each region (panels a$-$f). The BH is of spin $a_s=0.99$.
The composition parameter is $\xi=1.0$ or $\ep$ flow.}
\label{fig:fig3}
\end{figure}

In Fig. \ref{fig:fig3}, we divide the ${\cal E}-\lambda$ parameter space into regions that will admit
single critical points and MCP and shocks for accretion flows around a Kerr BH of $a_s=0.99$.
In the region ABD$^{\prime}$ the angular momentum is low and solutions in this domain will possess a single
outer-type critical point (typical solution Mach number $M=v/a$ versus $r$: panel a).
In the region ABHA, the angular momentum is higher and three sonic points ($\rci,~\rcm,~\rco$) appear,
although the flow still passes through $\rco$ on to the BH (typical solution in panel b).
The shaded region FGHF represents flow parameters for which steady-state shocks are possible. In other words,
matter with parameters from the region FGHF flows through $\rco$, becomes subsonic at the shock location
$\rsh$ and finally enters the BH through
$\rci$ (typical solution in panel c). Solutions from the region AHDEA are also characterized by three sonic points,
but since the entropy ${\dot {\cal M}}_{\rm i}$ of the inner critical point is higher than that of the
outer critical point, so the matter flows into the BH through $\rci$ (typical solution in panel d).
The parameters from the region above AEI have very high angular momentum such that it becomes transonic
close to the horizon producing a single inner-type critical point $\rci$. These solutions
are monotonic and smooth functions of $r$ (typical solution in panel e). Solutions from the region BDCB are bound
and do not produce global solution (typical solution in panel f). Parameters from the
rest of the region do not
produce transonic solutions that connect the horizon. The solid curves in panels above
are the physical solutions. The dotted
curves are solutions for the wind-type boundary condition and the dashed curves represent possible accretion
solutions but that are not chosen by the matter. 

\begin{figure}
	\centering
	\includegraphics[width=8.0cm]{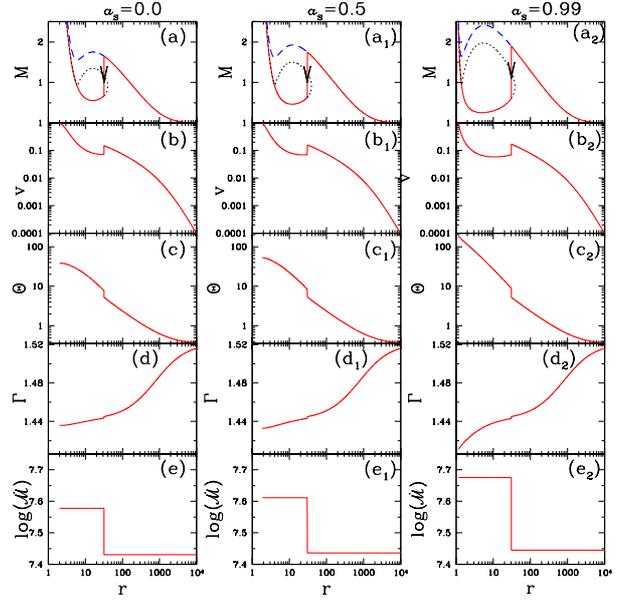}
	\caption{Accretion disc: $M$ (a, a$_1$, a$_2$), $v$ (b, b$_1$, b$_2$), 
		$\Theta$ (c, c$_1$, c$_2$), $\Gamma$ (d, d$_1$, d$_2$) and 
		${\dot {\cal M}}$ (e, e$_1$, e$_2$) are plotted 
		with $r$ along the equatorial plane. 
		Accretion on to a BH of $a_s=0.0$ is for $\lambda=3.05$ (a$-$e);
		those on to a BH of $a_s=0.5$ are for $\lambda=2.716$ (${\rm a}_1- {\rm e}_1$)
		and flows on to a BH with $a_s=0.99$ are plotted for $\lambda=2.01$ (${\rm a}_2 - {\rm e}_2$)
		For all plots ${\cal E}=1.001$ and $\xi=1.0$ and the shock location is at 
		$\rsh=30.40$. Arrows on vertical lines show the shock jump (a, a$_1$, a$_2$).
	}
	\label{fig:fig4}
\end{figure}

We now compare shocked accretion disc solutions
around BHs of three Kerr parameters $a_s=0$
(Figs. \ref{fig:fig4}a$-$e), $a_s=0.5$ (Figs. \ref{fig:fig4}${\rm a}_1-{\rm e}_1$), $a_s=0.99$ (Figs. \ref{fig:fig4}${\rm a}_2-{\rm e}_2$), where the flow angular momenta are $3.05$, $2.716$ and $2.01$, respectively. 
All the three cases presented above have same energy ${\cal E}=1.001$. The flow variables plotted as a function of $r$ are
the Mach number $M$ (Figs. \ref{fig:fig4}a$-$a$_2$), $v$ (Figs. \ref{fig:fig4}b$-$b$_2$),
$\Theta$ (Figs. \ref{fig:fig4}c$-$c$_2$), $\Gamma$ (Figs. \ref{fig:fig4}d$-$d$_2$)
and log$({\dot {\cal M}})$ (Figs. \ref{fig:fig4}e$-$e$_2$). The most interesting thing
is that the shock formed in all the three cases is
exactly at $\rsh=30.2$. 
In other words, accretion discs around BHs of different spin
may form shock at the same radial distance, if the angular momenta of the discs are different
even if ${\cal E}$ remains the same. 
Although $\rsh$ has the same value, the flow variables at any given $r$ are different for accretion discs around BHs of different $a_s$. As an example, the temperature of the PSD for BH of $a_s=0.99$ is much higher than that of
the flow on to a BH of $a_s=0$, so radiations computed from such flow would indeed be different, but if $\rsh$ is oscillating
there is a chance that it would be oscillating at the same frequency.

\subsection{All possible jet solutions}\label{subsec:jetcpaly}
In this section, we would like to discuss about all possible jet solutions: in other words,
we solve equations (\ref{dvdrj.eq}) and (\ref{dthj.eq}) with the help of equation (\ref{jtcp.eq}),
for given values of $Z_\phi$ (equation \ref{vZp.eq}). Since ${\cal R}_{\rj}$ is a constant of motion, so
we express ${\cal R}_{\rj}$ in terms of critical point conditions (equation \ref{jtcp.eq}) and solve for 
$r_{\rj \cm}$: we will obtain all possible sonic points. In addition, once we specify $Z_\phi$, the jet
cross-section
is specified and the entire jet solution can be obtained. In this section, ${\cal R}_{\rj}$
and $Z_\phi$ are supplied as free parameters, in order to find all possible solutions.
\begin{figure}
	\centering
	\includegraphics[width=8.0cm]{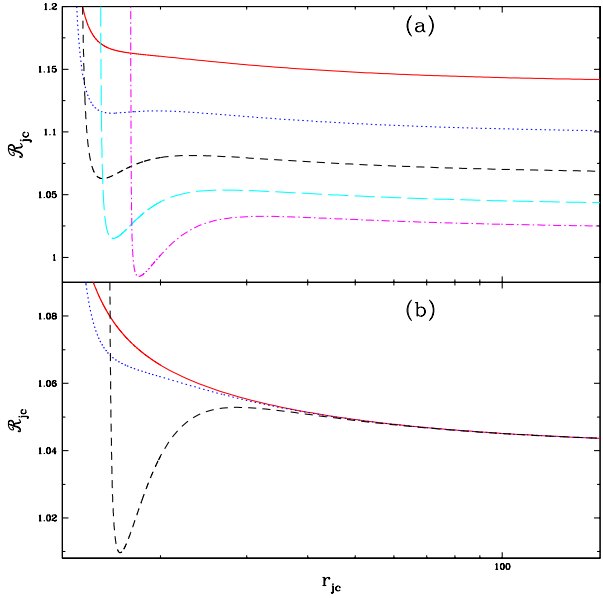}
	\caption{Jet: ${\cal R}_{\rj{\cm}}$ as a function of critical points $r_{\rj \cm}$
		for a given BH spin $a_s=0.9$ (a) and for a given flow angular momentum $\lambda_{\rj}=2$ (b).
		(a) Each curve represents $\lambda_{\rj}=3.5$ (solid, red), $\lambda_{\rj}=3.0$
		(dotted, blue), $\lambda_{\rj}=2.5$
		(dashed, black), $\lambda_{\rj}=2.0$ (long$-$dashed, cyan)
		and $\lambda_{\rj}=1.5$ (dash$-$dotted, magenta).
		(b) Each curve represents $a_s=0.0$ (solid, red), $a_s=0.5$ (dotted, blue)
		and $a_s=0.99$ (dashed, black).
		All the figures are plotted for the same von Zeipel parameter, $Z_{\phi}=10$ and $n_{\cm}=0.5$.
	}
	\label{fig:fig5}
\end{figure}

In Fig. (\ref{fig:fig5}a), we plot the jet energy parameter ${\cal R}_{\rj \cm}$ as a function of $r_{\rj \cm}$
for flows with $\lambda_{\rj}=3.5$ (solid, red), $\lambda_{\rj}=3.0$ (dotted, blue), $\lambda_{\rj}=2.5$
(dashed, black), $\lambda_{\rj}=2.0$ (long$-$dashed, cyan)
and $\lambda_{\rj}=1.5$ (dash$-$dotted, magenta) around a BH with spin $a_s=0.9$.
Here, ${\cal R}_{\rj \cm}={\cal R}_{\rj}|_{r_{\rj \cm}}$ and ${\dot {\cal M}}_{\rj \cm}=
{\dot {\cal M}}_{\rj}|_{r_{\rj \cm}}$.
It is clear from the above that jets with two different
$\lambda_{\rj}$ will have the same ${\cal R}_{\rj}$ and $r_{\rj \cm}$ at the crossing points. However, the entropy
(or ${\dot {\cal M}}_{o\cm}$) of the two flows with different $\lambda$ at the crossing points of Fig. (\ref{fig:fig5}a) are not the same. Another curious aspect is that the jet energy parameter
${\cal R}_{\rj\cm}>1$ even for $r_{\rj \cm} \rightarrow \infty$. Although this is counter-intuitive,
this arises due to the property of VZS, where not only $\lambda_{\rj}$ remains
constant but $\gamma_{\phi \rj}$ also remains constant: therefore,
${\cal R}_{\rj \cm} \rightarrow 
\left(\gamma_{\phi \rj}|_{\rm r_{\rb}}\right)^{1-2\beta}$ as $r_{\rj \cm}\rightarrow \infty$.
Moreover, in direct contrast with accretion critical point properties,
except at the crossing points of the curves, ${\cal R}_{\rj \cm}$ is generally higher for jets with
higher $\lambda_{\rj}$.
For example, curve for $\lambda_{\rj}=3.5$ is of higher ${\cal R}_{\rj \cm}$
than for $\lambda_{\rj}=3.0$ and so on.

In Fig. (\ref{fig:fig5}b), we once again plot ${\cal R}_{\rj\cm}$ as a function of $r_{\rj \cm}$,
but this time for the same flow $\lambda=2$, but falling on to BHs of different spin $a_s=0.0$
(solid, red), $\lambda=0.5$ (dotted, blue) and $a_s=0.99$ (dashed, black). There are no MCP
in jets around BHs of lower spin, but space$-$time around a rotating BH of
$a_s=0.99$ ensures multiple sonic points
in jets.

\begin{figure}
	\centering
	\includegraphics[width=8.0cm]{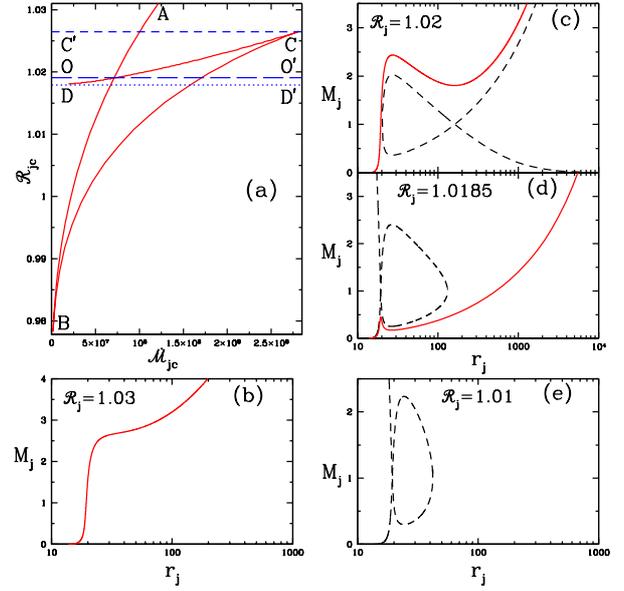}
	\caption{Jet: Variation of (a) ${\cal R}_{\rj\cm}$ with ($\dot{\cal M}_{\rj\cm}$) for $a_s=0.99$ (solid, red).
		Jet Mach number ($M_{\rj}$) with jet stream-line $r_{\rj}$ for different jet energy
		(b) ${\cal R}_{\rj}=1.03$, (c) ${\cal R}_{\rj}=1.02$, (d)
		${\cal R}_{\rj}=1.0185$ and (e) ${\cal R}_{\rj}=1.01$. Each solution is plotted for $Z_{\phi}=10, n_{\cm}=0.5$ and $\lambda_{\rj}=2.0$. 
		Solid and dashed curves represents global and closed jet solutions, respectively.  
	}
	\label{fig:fig6}
\end{figure}

Figure (\ref{fig:fig6}a) is the famous kite-tail relation between ${\cal R}_{\rj \cm}$ and
${\dot {\cal M}}_{\cm}$ for a given value of angular momentum, i. e.,
$\lambda_{\rj}=2$.
The curve CD represents the values of ${\cal R}_{\rj \cm}$ and ${\dot {\cal M}}_{\rj \cm}$
for all possible
$\rjco$. Similarly, curves AB and BC represent the values of ${\cal R}_{\rj \cm}$ and ${\dot {\cal M}}_{\rj \cm}$
for all possible $\rjci$ and $\rjcm$, respectively. Here,
the inner, middle and outer jet sonic points are represented by
$\rjci$, $\rjcm$ and $\rjco$. In Figs. (\ref{fig:fig6}b-e), we plot the related jet solutions,
i. e., jet Mach number $M_{\rj}$ as a function of $r_{\rj}$. 
Horizontal line OO$^{\prime}$ (long$-$dashed) passes through
the intersection of AB and CD at ${\cal R}_{\rj \cm}=1.0193$, and
CC$^{\prime}$ (dashed) corresponds to
highest ${\cal R}_{\rj \cm}~(=1.0265)$ for which MCP are possible.
And DD$^{\prime}$ is
lowest ${\cal R}_{\rj \cm}~(=1.0181)$ for which three sonic points are possible.
For ${\cal R}_{\rj \cm}<1.0181$ only
closed jet solutions are possible with only two sonic points $\rjci$ and $\rjcm$,
but no $\rjco$. 
And therefore, in Fig. (\ref{fig:fig6}b), there is one inner-type critical point since ${\cal R}_{\rj}=1.03$
lies above the CC$^{\prime}$ in ${\cal R}_{\rj \cm}$---${\dot {\cal M}}_{\cm}$ space. The energy parameter
of the jet in Fig. (\ref{fig:fig6}c), is ${\cal R}_{\rj}=1.02$ which is in between CC$^{\prime}$ and
OO$^{\prime}$. In this domain, both $\rjci$ and $\rjco$ are possible and the entropy of $\rjco$ is higher,
but the solution (dashed) through $\rjco$ is a closed one (dashed),
while the one through $\rjci$ is global (solid). In Fig. (\ref{fig:fig6}d), ${\cal R}_{\rj}=1.0185$ lies between
OO$^\prime$ and DD$^{\prime}$, and $\rjci$, $\rjco$ still form, but now the global solution
(solid) is through $\rjco$ and the closed solution is through $\rjci$. It is clear from Fig. (\ref{fig:fig6}a),
that the inner critical point has higher entropy for
solutions with ${\cal R}_{\rj}$ between OO$^\prime$ and DD$^{\prime}$. In Fig. (\ref{fig:fig6}e),
${\cal R}_{\rj}=1.01$ below DD$^{\prime}$ possesses a X$-$type $\rjci$ and O$-$type $\rjcm$.
The jet flows out through $\rjci$ but turns back around the O$-$type sonic points, i. e. there
is no global solution, only a closed one.

\begin{figure}
	\centering
	\includegraphics[width=7.0cm]{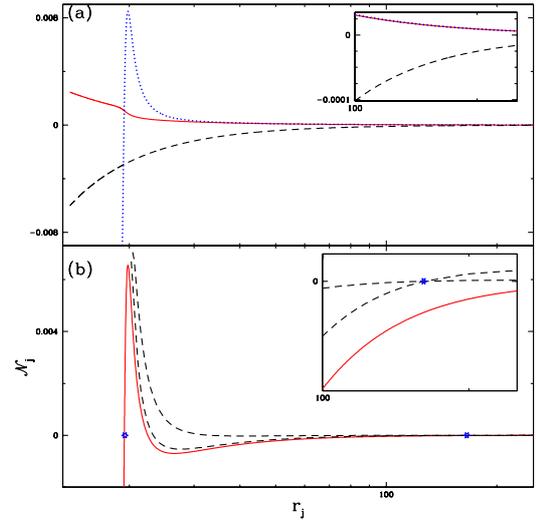}
	\caption{Jet: various terms of ${\cal N}_{\rj}$ like (a) ${a_{\rj}^2}{({\cal A}_{\rj})^{-1}}(d{\cal A}_{\rj})/(dr_{\rj})$ (solid, red),
		$-{a_{\rj}^2}{h_p}^{-1}(dh_p)/(dr_{\rj})$ (dotted, blue) and
		$-X_{\rm g}$ (dashed, black) are plotted with $r_{\rj}$.
		(b) ${\cal N}_{\rj}$ for the global solution through $\rjci$ (solid, red)
		and the closed solution through $\rjco$ (dashed).
		Asterisk shows the location of two X-type sonic points.
		Other parameters $Z_{\phi}=10, n_{\cm}=0.5$ and $\lambda_{\rj}=2.0$.
		Same jet solution as in Fig. (\ref{fig:fig6}c).
	}
	\label{fig:fig7}
\end{figure}

In Fig. (\ref{fig:fig6}c), we observe that the jet solution admits multiple critical points.
In the case of accretion disc, the gravity ensures one critical point, but if the matter is rotating
then centrifugal term modifies the gravitational interaction and forms multiple sonic or critical points.
However, in jets the effect of centrifugal term along the streamline is quite small. 
 In Fig. (\ref{fig:fig7}a),
we plot various terms of the numerator ${\cal N}_{\rj}$ of equation (\ref{dvdrj.eq}) in order to understand why we have MCP, for the same jet parameters as considered
in Fig. (\ref{fig:fig6}c). In Fig. (\ref{fig:fig7}b), we plot the ${\cal N}_{\rj}$ for all the branches
of the jet solution for $R_{\rj}=1.02,~\lambda=2.0$. From equation (\ref{dvdrj.eq}) it is clear that
the jet will accelerate ($dv_{\rj}/dr_{\rj}>0$) if ${\cal N}_{\rj}<0$
in the subsonic ($v_{\rj}<a_{\rj}$) regime and
changes sign in the supersonic regime
($v_{\rj}>a_{\rj}$).
${\cal N}_{\rj}$ becomes zero at the critical point. 
In Fig. (\ref{fig:fig7}a), we plot
the three components of ${\cal N}_{\rj}$, \ie~ ${\cal N}_{\rj 1}={a_{\rj}^2}{({\cal A}_{\rj})^{-1}}(d{\cal A}_{\rj})/(dr_{\rj})$ (solid, red),
${\cal N}_{\rj 2}=-{a_{\rj}^2}{h_p}^{-1}(dh_p)/(dr_{\rj})$ (dotted, blue) and
${\cal N}_{\rj 3}=-X_{\rm g}$ (dashed, black), as a function of $r_{\rj}$. ${\cal N}_{\rj 1}$,
the coupling between thermal
and the flow geometry term, 
is positive definite and ${\cal N}_{\rj 3}$, the gravity term, is negative definite. But it is ${\cal N}_{\rj 2}$, the thermal and metric term, which 
flips sign, that determines the formation of multiple sonic points and eventually shock. Near the base, where the jet is subsonic, Fig. (\ref{fig:fig7}a) shows ${\cal N}_{\rj 2}<0$ and
dominates both ${\cal N}_{\rj 1}$ and ${\cal N}_{\rj 3}$.
Therefore,
${\cal N}_{\rj}<0$ near the base (Fig. \ref{fig:fig7}b) and this ensures that the jet accelerates in subsonic region.
At $r_{\rj ci}$ or the inner critical point, ${\cal N}_{\rj}$ becomes zero (inner star in Fig. \ref{fig:fig7}b),
as ${\cal N}_{\rj 2}$ becomes positive enough to negate gravity. Further out, ${\cal N}_{\rj 2}$
remains positive and increases up to a short distance, so the jet continues to accelerate. However, ${\cal N}_{\rj 2}$
reaches its maximum and starts to decrease rapidly. At few$\times 10~\rg$, ${\cal N}_{\rj 2}
\approx{\cal N}_{\rj 1}\ll |{\cal N}_{\rj 3}|$, which implies that ${\cal N}_{\rj}<0$ at large $r_{\rj}$
(solid, red in Fig. \ref{fig:fig7}b).
Therefore, the global solution of the jet through $r_{\rj ci}$ decelerates, after it becomes supersonic.
This resistance to the supersonic flow creates the possibility for
the jet to pass through the second saddle$-$type sonic point.
And since the dynamics is forcing the jet to choose another solution, the entropy of the outer critical point is also high. Needless to say, in order to flow out through $\rjco$, the jet has to generate the right amount
of entropy via a shock. We plot the ${\cal N}_{\rj}$ of the solution through the outer sonic point, and one can see that it is positive beyond the outer sonic point (dashed in Fig. \ref{fig:fig7}b). The inset panels in Fig. (\ref{fig:fig7}a \& b) are plotted to zoom around the
outer sonic point (asterisk), to show the different branches of the solution.

\subsection{Self-consistent accretion-jet solution}\label{subsec : accjet}
\begin{figure}
 \centering
 \includegraphics[width=8.0cm]{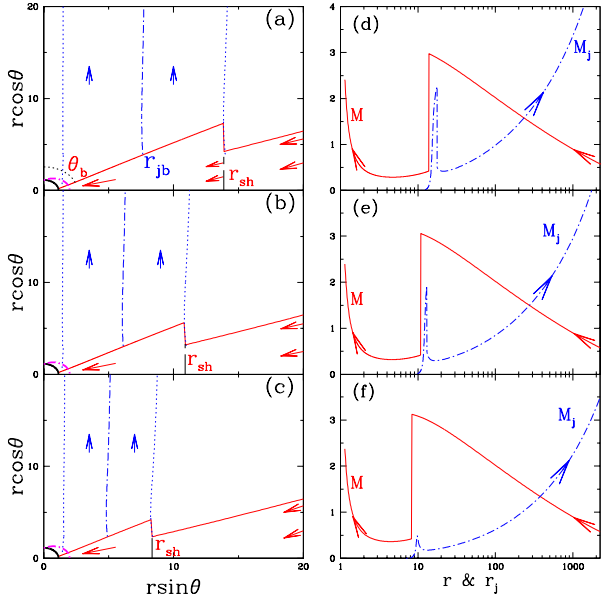}
 \caption{Self-consistent accretion-jet solutions: flow boundary (a, b, c) of accretion disc (solid, red),
 	jet streamline (dash$-$dotted, blue), jet boundary (dotted, blue), horizon (solid quarter circle),
 	ergosphere (dash$-$dotted quarter ellipse), the radius of the jet base $r_{\rj b}$ and its polar
 	angle $\theta_{\rj b}$ are shown. Accretion Mach number $M$ (solid, red) and jet Mach number
 	$M_{\rj}$ (dash$-$dotted blue) with radial distance are plotted (d, e, f). The right-hand panels
 	correspond to the accretion-jet solutions and the left-hand panels correspond to related
 	geometry.
 Each pair of right- and left-hand panels are plotted for $\lambda=2$ (a, d), $\lambda=1.99$
 (b, e) and $\lambda=1.98$ (c, f). Accretion shocks are produced at $\rsh=13.8308$ (a, d), $\rsh=10.8179$
(b, e) and $\rsh=8.2275$ (c, f). 
The arrows show flow direction.
For all the panels ${\cal E}=1.0001, \xi=1.0$ and $a_s=0.99$.}
 \label{fig:fig8}
\end{figure}

In this section, we present accretion$-$ejection solutions.
We connect Sections \ref{subsec:accret} and \ref{subsec:jetcpaly}
where the inner boundary condition of the jet is obtained from the PSD,
i.e. we follow steps (i)---(v) of
Section \ref{sec:method} in totality.
In other words, we now obtain simultaneous accretion-jet solutions by supplying
accretion disc parameters like ${\cal E}, \lambda$, flow composition $\xi$ and BH spin $a_s$.
In Figs (\ref{fig:fig8}a-f), we plot accretion$-$ejection solutions for same disc specific energy
${\cal E}=1.0001$, but different angular momenta $\lambda=2.0$ (Figs \ref{fig:fig8}a, d),
$\lambda=1.99$ (Figs \ref{fig:fig8}b, e) and $\lambda=1.98$ (Figs \ref{fig:fig8}c, f). In the left-hand panels,
we plot the disc$-$jet geometry (Figs \ref{fig:fig8}a$-$c).
In the right-hand panels (Figs \ref{fig:fig8}d$-$f),
we plot accretion
disc Mach number $M$ (solid, red) and jet Mach number $M_{\rj}$ with distance. $M$ is evaluated on the equatorial plane
while $M_{\rj}$ along the jet stream line (dash$-$dotted on the left-hand panels).
All the accretion solutions presented here harbour shock, while the top two panels on the right
(Figs. \ref{fig:fig8}d \& e) show shocks in
jets too. The shock in accretion decreases with decreasing $\lambda$, and so does the
jet cross-section. The jet starts from the surface of the PSD with very low velocity
but either expands briskly to become transonic within a short distance and eventually gets shocked, or
slowly becomes transonic at a large distance from the central object.
In this particular case, the shock strength of the jet is slightly less than that of the accretion disc.

\begin{figure}
 \centering
 \includegraphics[width=8.0cm]{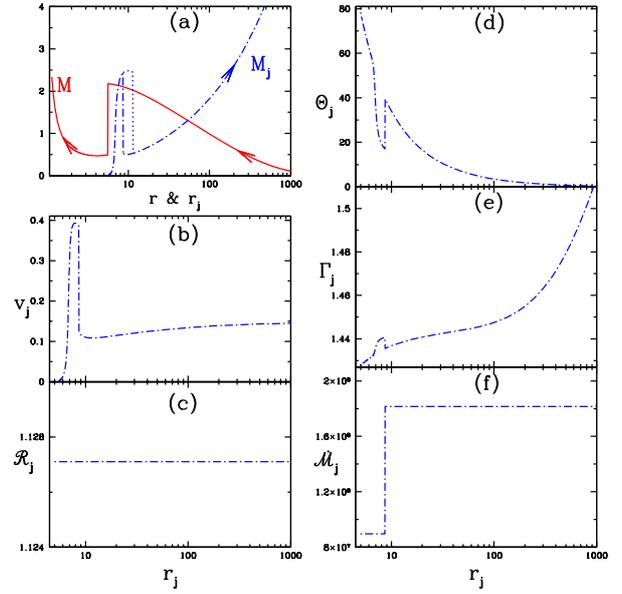}
 \caption{Variations of (a) accretion Mach number $M$ (solid ,red) and jet Mach number
 $M_{\rj}$ (dash$-$dotted,blue) 
 with radial distance on disc equatorial plane $r$ and jet streamline above the
 equatorial plane, $r_{\rj}$, respectively:
 Variation of jet flow variables $v_{\rj}$ (b), ${\cal R}_{\rj}$ (c), $\Theta_{\rj}$ (d),
 $\Gamma_{\rj}$ (e)
 and $\dot{\cal M}_{\rj}$ (f) with $r_{\rj}$. Arrow represents the direction of the flow.
 Disc parameters are ${\cal E}=1.002,~ \lambda=1.946, \xi=1.0$ and $a_s=0.99$. The accretion 
 shock location is at $\rsh=5.9827$. The relative mass outflow rate is $\rmdot=0.061378$. 
 }
 \label{fig:fig9}
\end{figure}

In Fig. (\ref{fig:fig9}a), we plot simultaneous accretion and jet Mach numbers $M$ and $M_{\rj}$, respectively,
where there are shocks both in accretion and the jet. In the rest of the figure we plot the jet flow variables
like jet three-velocity $v_{\rj}$ (Fig. \ref{fig:fig9}b), jet energy parameter $R_{\rj}$ (Fig. \ref{fig:fig9}c),
$\Theta_{\rj}$ (Fig. \ref{fig:fig9}d), $\Gamma_{\rj}$ (Fig. \ref{fig:fig9}e) and ${\dot {\cal M}}_{\rj}$
(Fig. \ref{fig:fig9}e), for the given accretion disc parameters ${\cal E}=1.002,~ \lambda=1.946, \xi=1.0$
around a BH of $a_s=0.99$. The mass outflow rate obtained is $\rmdot =0.061378$. The accretion shock is at $\rsh=5.9827$ on the equatorial plane, while
the shock in the jet is at $r_{\rj sh}=8.597$ at $\theta_{\rj sh}=24.64^{\circ}$. There is another shock location for the
jet that is slightly further away (at a distance of $11.448$, dotted vertical line),
but can be shown to be unstable \citep{n94,n96}.
Interestingly, the shock in accretion is at a shorter distance than the shock in jet: therefore, the
inner part of the disc$-$jet system has a complicated structure. The jet starts from the surface of the PSD
with almost negligible velocity, is accelerated within a few gravitational radii to about $\sim 0.4$,
and then jumps down to about $\sim 0.1$, and finally its terminal velocity is $v_{\rj \infty}\sim 0.146$.
The jet shock is quite strong (compression ratio $\sim 4$), and should be a good site for particle
acceleration. A major feature of shocks in jets is that the base of the jet
is much hotter than the pre-shock portion as well as the post-shock region of the jet.
In contrast, PSD is the hottest part of the accretion disc. 
The jet energy parameter
remains constant (Fig. \ref{fig:fig9}c), although the entropy jumps at the shock (Fig. \ref{fig:fig9}e).
  
 \begin{figure}
 \centering
 \includegraphics[width=8.0cm]{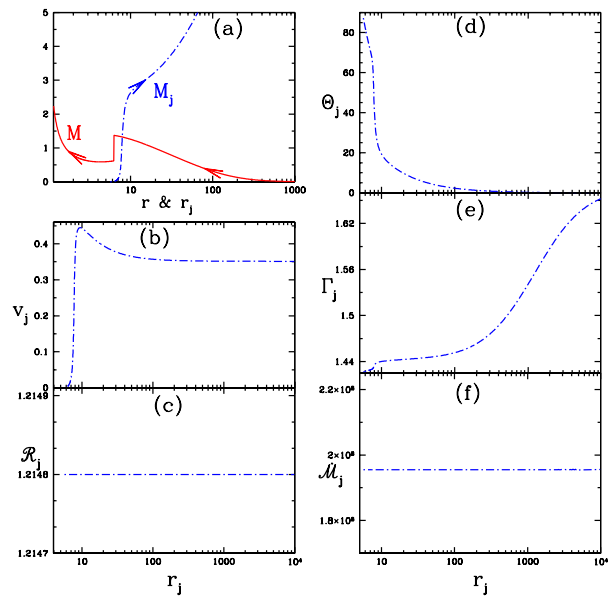}
 \caption{Variations of (a) $M$ (solid, red) and $M_{\rj}$ (dash$-$dotted, blue) with  
 $r$ and $r_{\rj}$, respectively. Variation of $v_{\rj}$ (b),
 ${\cal R}_{\rj}$ (c), $\Theta_{\rj}$ (d), $\Gamma_{\rj}$ (e) and 
 $\dot{\cal M}_{\rj}$ (f) with jet streamline $r_{\rj}$. Arrow represents the direction of the flow.
 Disc parameters are ${\cal E}=1.0105, \lambda=1.903, \xi=1.0$ and $a_s=0.99$ and accretion  
 shock formed at $\rsh=6.2941$. The $\rmdot=0.054495$.
 }
 \label{fig:fig10}
\end{figure} 
 In Fig.(\ref{fig:fig10}a), we plot accretion disc solutions $M$ (solid, red) and the corresponding jet
 $M_{\rj}$ (dash$-$dotted, blue) for disc parameters are ${\cal E}=1.0105, \lambda=1.903, \xi=1.0$ and $a_s=0.99$.
 The jet velocity $v_{\rj}$ (Fig.\ref{fig:fig10}b), ${\cal R}_{\rj}$ (Fig.\ref{fig:fig10}c), $\Theta_{\rj}$
 (Fig.\ref{fig:fig10}d), $\Gamma_{\rj}$ (Fig.\ref{fig:fig10}e) and ${\dot {\cal M}}_{\rj}$ (Fig.\ref{fig:fig10}f).
 This particular set of disc parameters launches a jet that is free from shock. In this case too, the jet launched by PSD starts
 with negligible velocity at the base, but rapidly accelerates to become transonic in the next few $\rg$ (Fig.\ref{fig:fig10}b),
 by converting thermal energy into kinetic energy (Fig.\ref{fig:fig10}d). Since this jet is shockfree, so
 in this case the entropy remains constant (Fig.\ref{fig:fig10}f) and like the previous case ${\cal R}_{\rj}$
 is constant too.
 Interestingly, as the jet is rapidly accelerated
 by thermal gradient term, it reaches a speed $v_{\rj}>0.44$, but eventually the thermal gradient term
 exhausts itself, and the jet settles to a lesser terminal speed of $v_{\rj \infty}\sim 0.3525$.
 The temperature drops sharply in the rapid acceleration phase, but at a lesser rate in the region where the jet velocity is approaching an asymptotic value. The jet temperature indeed approaches non-relativistic
 values at large distances (Fig.\ref{fig:fig10}d). This is also shown in the $\Gamma_{\rj}$ variation too
 (Fig.\ref{fig:fig10}e). 
  
\begin{figure}
 \centering
 \includegraphics[width=8.0cm]{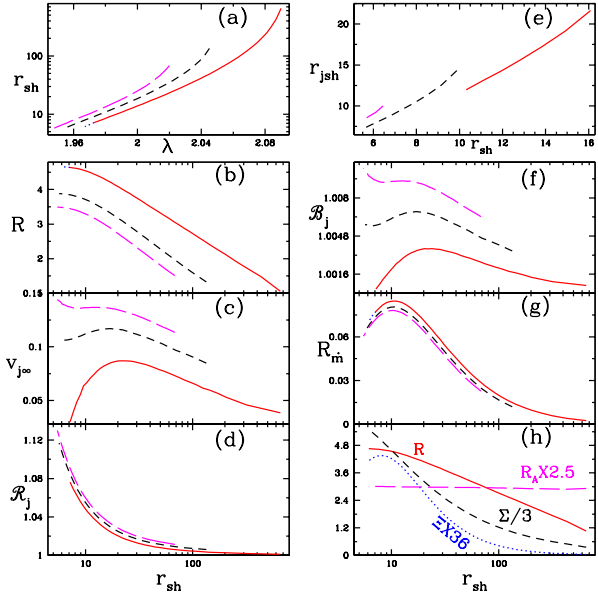}
 \caption{Variations of (a) $\rsh$ with $\lambda$, and (b) $R$, (c)  $v_{\rj\infty}$, (d) ${\cal R}_{\rj}$,
 (e) $r_{\rj \rm sh}$, (f) ${\cal B}_{\rj}$ and (g) $R_{\dot m}$ with $\rsh$.  
 Each curve is for ${\cal E}=1.0001$ (solid, red), $1.001$ (dashed, black)
 and $1.0019$ (long-dashed, magenta). (h) Variation of $R$ (solid), $R_A$ (long-dashed),
 $\Sigma$ (dashed) and $\varXi$ (dotted) with $\rsh$ for ${\cal E}=1.0001$. For all the panels $\xi=1.0$ and $a_s=0.99$.
 }
 \label{fig:fig11}
\end{figure}
In Fig. ({\ref{fig:fig11}a}), we plot the disc shock location $\rsh$ with disc angular momentum $\lambda$.
Each curve is for ${\cal E}=1.0001$ (solid, red), $1.001$ (dashed, black)
 and $1.0019$ (long-dashed, magenta). All the curves have fixed values of $a_s=0.99$ and $\xi=1.0$. 
The shock $\rsh$ increases with increasing $\lambda$ at a given ${\cal E}$, as well as
increasing ${\cal E}$ at a given $\lambda$. 
We plot the related compression ratio $R$ (Fig.{\ref{fig:fig11}}b), jet terminal speed $v_{\rj\infty}$ (Fig.{\ref{fig:fig11}}c), jet energy parameter ${\cal R}_{\rj}$ (Fig.\ref{fig:fig11}d), the jet shock location $r_{\rj sh}$(Fig.{\ref{fig:fig11}}e),
effective jet specific energy ${\cal B}_{\rj}$ (Fig.{\ref{fig:fig11}}f)
and $\rmdot$ (Fig.{\ref{fig:fig11}}g) with the accretion shock location $\rsh$.
The compression ratio $R$ of the shock in accretion disc increases with the decreasing $\rsh$
as is shown in Fig. (\ref{fig:fig11}b). Compression ratio decreases for weak shock, and for accretion disc shocks, shock becomes weaker as $\rsh$ increases. However, only shock compression
does not power the jet, and it shows that $v_\infty$ has a rather complicated dependence on $R$,
where the terminal speed has a clear maximum for ${\cal E}=1.0001$ (solid, red), although $R$
decreases with increasing $\rsh$. Moreover, the jet terminal speed is generally higher
for higher energies like ${\cal E}=1.00055$ (dashed, black) and ${\cal E}=1.001$ (long-dashed, magenta),
although the compression ratio of the accretion shock for such energies is clearly low.
Here, $v_{\rj \infty}$ is not a monotonic function of $\rsh$, but the jet energy parameter ${\cal R}_{\rj}$
decreases monotonically with increasing $\rsh$ (or decreasing $R$). ${\cal R}_{\rj}$ is also higher
for accretion discs of higher energy. The shock in jet is only formed in a limited range
as is shown in Fig. (\ref{fig:fig11}e). The shock in jet, {\ie} $r_{\rj sh}$, increases as disc energy increases:
however, the range of steady shock in jet decreases, if the energy of the disc is higher.
The relative Bernoulli parameter has similar functional dependence on $\rsh$ as $v_{\rj \infty}$, which
is not surprising since at $r\rightarrow \infty$, ${\cal B}\rightarrow (1-v^2_{\rj \infty})^{-1/2}$.
It is worth noting that the relative mass outflow rate or $\rmdot$ does not follow 
the functional dependence of $\vjinf$ with $\rsh$. $\rmdot$ generally increases with decreasing
$\rsh$, but also decreases for very small value of $\rsh$, maximizing at some intermediate values of
$\rsh$. This is because $\vjinf$ only depends on the ${\cal B}_{\rj}$, but $\rmdot$ depends on many factors
as shown in Section \ref{subsec:shockeqn}). In order to understand how $\rmdot$ depends on $\rsh$, in the next panel
we analyse only the case presented in this figure, {\ie} the case for ${\cal E}=1.0001$ (solid, red in Figs. \ref{fig:fig11}a-g). In Fig. (\ref{fig:fig11}h), we plot the accretion compression ratio
$R$ (solid, red), the ratio of jet cross-section at the base to the PSD, {\ie} $R_A$ (long-dashed, magenta),
density contrast across accretion shock, {\ie} $\Sigma$ (dashed, black) and $\varXi$ (dotted, blue) which is the ratio of the relativistic mass flux
at the jet base to that of the pre-shock accretion disc: it gives a measure of the upward thrust
that the shock generates. From Section \ref{subsec:shockeqn}, it is clear that
if $\Sigma$ increases together with the decrease of $R_A$, $R$ and $\varXi$, then $\rmdot$ will decrease too
and vice versa.
As $\rsh$ decreases by changing $\lambda$ for ${\cal E}=1.0001$, $\Sigma$ monotonically increases,
$R$ increases but tends to taper off for small $\rsh$, $\varXi$ increases but starts to dip for small
$\rsh$ and $R_A$ do not change much. Therefore, $\rmdot$ generally increases for decreasing $\rsh$,
but dips for small values of $\rsh$, which points to the fact that the combined effect of $R$ and $\varXi$ 
somewhat negates the effect of $\Sigma$ in determining the value of $\rmdot$.

\begin{figure}
 \centering
 \includegraphics[width=8.0cm]{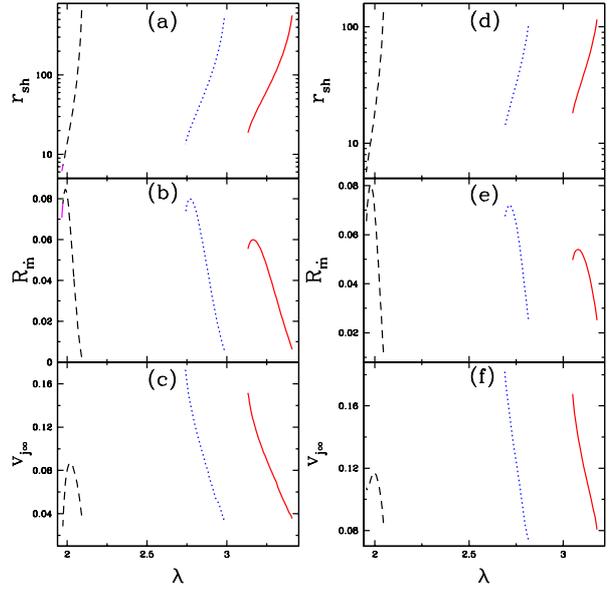}
 \caption{Variations of $\rsh$ (a, d), $\rmdot$ (b, e) and $v_{\rj\infty}$ (c, f) with ${\lambda}$
 are plotted
for different BHs of Kerr parameter $a_s=0.0$ (solid, red), $0.5$ (dotted, blue)
and $0.99$ (dashed, black). 
The disc parameters are ${\cal E}=1.0001$ (a, b, c) and ${\cal E}=1.001$ (d, e, f).
The flow composition for all the plots is $\xi=1.0$.
 }
 \label{fig:fig12}
\end{figure}
In Fig.(\ref{fig:fig12}), we study the dependence of accretion shock $\rsh$ (Figs. \ref{fig:fig12} a, d), the relative outflow rate $\rmdot$ (Figs. \ref{fig:fig12} b, e) and the jet terminal speed $\vjinf$
(Figs. \ref{fig:fig12} c, f)
with the angular momentum $\lambda$
of the accretion disc. Each curve represents BHs of spin
$a_s=0.0$ (solid, red), $0.5$ (dotted, blue)
and $0.99$ (dashed, black). The figures on the left-hand panels (Figs. \ref{fig:fig12} a$-$c)
are plotted for disc energy ${\cal E}=1.0001$ and those on the right-hand panels
(Figs. \ref{fig:fig12} d$-$f)
are plotted for ${\cal E}=1.001$. In the top panels, we show how disc property
like $\rsh$ depends
on disc parameter, while in the middle and bottom panels, we show how jet properties
like $\rmdot$ and $\vjinf$ depend on the disc parameter $\lambda$. The accretion shock increases
with increasing disc $\lambda$ for BHs with all possible spins. It is worth noting that
accretion shocks may form even for low $\lambda$ values, if BH spin is high enough
(Figs. \ref{fig:fig12} a and d). Figs (\ref{fig:fig12}a) and (d) vindicate the conclusions
of Figs (\ref{fig:fig4}a$_1$) and (a$_2$),
that accretion shock may occur at the same location, for discs of different $\lambda$
depending on the spin of the BH.
The mass outflow rate from an accretion disc around a highly spinning
BH is higher than those around BHs of smaller $a_s$ (Figs. \ref{fig:fig12} b, e), but $\rmdot$ peaks
at intermediate $\lambda$ for BHs of any given spin.
However, $\vjinf$ for a given disc energy is not necessarily
higher for discs around highly spinning BHs, although it tends to increase with the flow energy
(Figs. \ref{fig:fig12} c, f). While, for jets around low-$a_s$ BH, $\vjinf$ increases monotonically as $\lambda$ is decreased,
but for highly spinning BH, $\vjinf$ maximizes at intermediate values of $\lambda$.

\begin{figure}
 \centering
 \includegraphics[width=9.0cm]{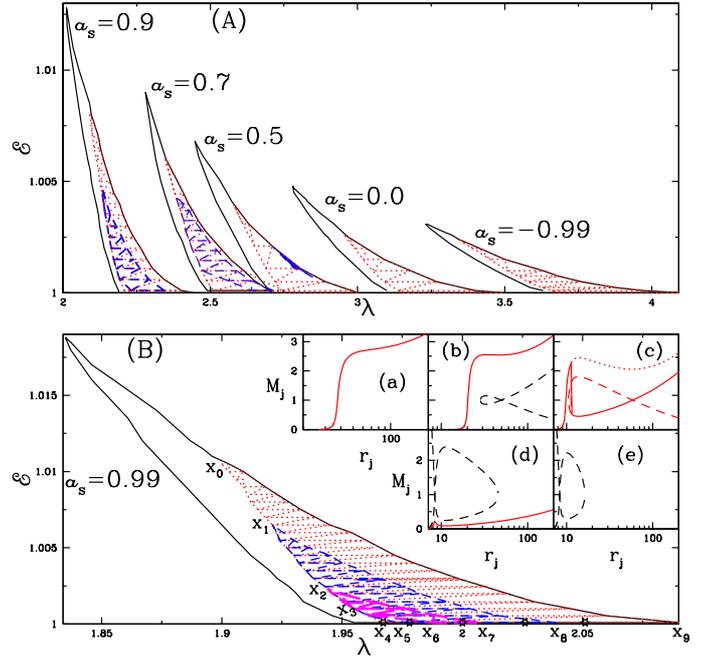}
 \caption{(A) Comparing ${\cal E}-\lambda$ accretion shock parameter space for $a_s=-0.99, 0.0, 0.5, 0.7$ and $ 0.9$. (B) $a_s=0.99$: ${\cal E}-\lambda$ (solid black) space for accretion shock.
Region x$_0$x$_4$x$_9$x$_0$ (dotted, red curve) represents accretion shock when mass-loss is computed. The region x$_1$x$_4$x$_8$x$_1$ represents ${\cal E}-\lambda$
for which jet possesses multiple (two or three) critical points. The region x$_1$x$_5$x$_8$x$_1$
(dashed, blue curve) represents the accretion ${\cal E},~\lambda$ for which
 the jet has three critical points. The region x$_2$x$_6$x$_7$x$_2$ (long-dashed, magenta curve)  represents ${\cal E},~\lambda$ for which jet harbours shock. 
 Inset: jet Mach number ($M_{\rj}$) with $r_{\rj}$ for different (a) $\lambda=2.05$ from
 region x$_0$x$_1$x$_8$x$_9$x$_0$;
 (b) $\lambda=2.025$ from x$_1$x$_5$x$_8$x$_1$; (c) $\lambda=2.0$ from x$_2$x$_6$x$_7$x$_2$, (d) $\lambda=1.98$ from x$_3$x$_5$x$_6$x$_3$ and (e) $\lambda=1.97$ from x$_3$x$_4$x$_5$x$_3$. 
 The accretion disc parameters are ${\cal E}=1.0001, \xi=1.0$. 
 }
 \label{fig:fig13}
\end{figure}
In Fig.(\ref{fig:fig13}), we present the ${\cal E}$---$\lambda$ parameter space of shock
in the accretion disc. 
The region bounded by the solid (black) curve represents ${\cal E}$ and $\lambda$ of the accretion disc
that will 
experience steady shock transition if we ignore the formation of jet. If 
mass-loss is taken into account, then the region x$_0$x$_4$x$_9$x$_0$ bounded by the dotted (red) curve
represents the parameter space for steady accretion shock that launches jet. The region x$_1$x$_4$x$_8$x$_1$
represent the accretion disc parameters for which jets would harbour multiple (two or three) critical points. And the region x$_1$x$_5$x$_8$x$_2$ bounded by dashed (blue) curve represents accretion disc parameters for which the jet possess three critical points. And in the region x$_2$x$_6$x$_7$x$_2$ within
long$-$dashed (magenta) curve, jet launched from the disc undergoes steady shock
transition. In the rest of the accretion shock domain (solid, black curve), there is no jet possible. In Fig. (\ref{fig:fig13}A), we compare the shock parameter space for BHs with spin parameters
$a_s=-0.99,~0.0,~0.5,~0.7,~0.9$. In Fig. (\ref{fig:fig13}B), we plot the shock parameter space for $a_s=0.99$
and in the inset panels, we plot only the jet solution of the disc$-$jet system {\ie} $M_{\rj}$ as a function of $r_j$
for accretion disc parameters ${\cal E}=1.0001$ and $\lambda=2.05$ (Fig. \ref{fig:fig13}Ba), $\lambda=2.025$
(Fig. \ref{fig:fig13}Bb), $\lambda=2.0$ (Fig. \ref{fig:fig13}Bc), $\lambda=1.98$ (Fig. \ref{fig:fig13}Bd)
and $\lambda=1.97$ (Fig. \ref{fig:fig13}Be). The disc parameters chosen
for $a_s=0.99$ are marked as stars in ${\cal E}-\lambda$ space (Fig. \ref{fig:fig13}B).
The angular momentum value of x$_4$ is $\lambda_{{\rm x}_4}=1.967$, x$_5$ is
$\lambda_{{\rm x}_5}=1.972$, x$_6$ is $\lambda_{{\rm x}_6}=1.988$, x$_7$ is $\lambda_{{\rm x}_7}=2.006$, x$_8$ is $\lambda_{{\rm x}_8}=2.039$ and x$_9$ is
$\lambda_{{\rm x}_9}=2.09$.
Each set of the disc parameters are chosen such that all possible jet solutions can be obtained. It is interesting to note that the parameter space for steady shocks shrinks, when self-consistent massloss is considered.
It means that there is a possibility of shock instability that is caused by the launching of the jets.   
Figure (\ref{fig:fig13}Be) shows that the resulting jet solution is actually a closed solution.
Although the accretion disc energy ${\cal E}>1$ and so is the jet Bernoulli parameter
${\cal E}_{\rj}$ (see equation \ref{Berj.eq}), the effective Bernoulli parameter
for the jet ${\cal B}_{\rj}<1$. Therefore, there is no global jet solution, for this ${\cal E}$
and $\lambda$. In other words, we predict from theoretical considerations
that even if there is an accretion shock, the PSD may not always launch a jet. 
It may be noted that in Figs (\ref{fig:fig12}c) and (f), the jet terminal speed obtained for low to moderate disc energies
decreases with the increase of $a_s$. However, from Figs (\ref{fig:fig13} A) and (B), we observe that the maximum value of energy parameter (of Fig. \ref{fig:fig13}) for which steady
accretion shock is possible increases with $a_s$, which might launch jets
with higher terminal speeds. 
The various coordinates e.g. x$_0-{\rm x}_9$ in ${\cal E}-\lambda$ accretion shock parameter space of Fig. (\ref{fig:fig13}B) and corresponding jet properties are tabulated in Table \ref{tab:tab1}.
\begin{table}
\begin{tabular}{c c c c c c}
\hline \hline
Points & Accretion & Accretion & $v_{\rj \infty}$ & $R_{\dot m}$ & Shock  \\
in ${\cal E}-\lambda$ & ${\cal E}$ & $\lambda$&&&in jet \\
\hline
x$_0$&1.0105 &1.903 & 0.353 & 0.055& NO \\
\hline
x$_1$ & 1.0065 & 1.921 & 0.232 & 0.059& NO \\
\hline
x$_2$ & 1.002 &1.946 &0.146 &0.062 & YES \\
\hline
x$_3$&1.00035&1.964&Bound& NA & NA \\
 & & &Jet&& \\
 \hline
x$_4$&1.0001&1.967&Bound&NA&NA  \\
&&&Jet& & \\
\hline
x$_5$&1.0001&1.973&0.028&0.0774&NO \\
\hline
x$_6$&1.0001&1.988&0.0667&0.0845&YES \\
\hline
x$_7$&1.0001&2.006&0.0838&0.0782&YES \\
\hline
x$_8$&1.0001&2.039&0.08196&0.045&NO \\
\hline
x$_9$&1.0001&2.09&0.0384&0.0025&NO \\
\end{tabular}
\caption{The coordinates of locations x$_0-{\rm x}_9$ in the ${\cal E}-\lambda$ parameter space
of the accretion disc as in Fig.(\ref{fig:fig13}B) and the
measure of corresponding jet properties in terms of $v_{\rj \infty}$ and $R_{\dot m}$.}
\label{tab:tab1}
\end{table}

\begin{figure}
 \centering
 \includegraphics[width=8.0cm]{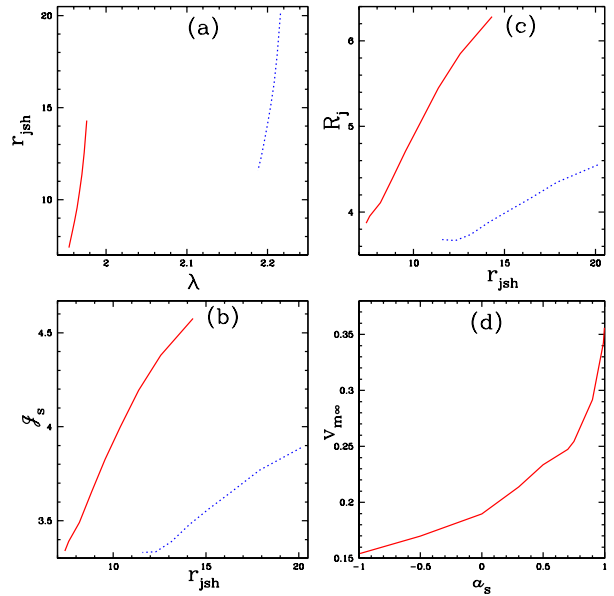}
 \caption{Variation of (a) jet shock location ($r_{\rj sh}$) with $\lambda$,
 (b) jet shock compression ratio $R_{\rj}$ and (c) jet shock strength ${\cal J}_{s}$ with
 $r_{\rj sh}$, where each curve is for $a_s=0.99$ (solid, red), 
 $0.9$ (dotted, blue) and all the plots are for ${\cal E}=1.001$. 
 (d) Maximum jet terminal velocity ($v_{\rm m \infty}$) with $a_s$.  }
 \label{fig:fig14}
\end{figure}

Fig. \ref{fig:fig14}(a) represents variation of jet shock location $r_{\rj sh}$
with the disc angular momentum $\lambda$, around two BHs of spin $a_s=0.99$ (solid, red) and
$a_s=0.9$ (dotted, blue). Corresponding jet compression ratio $R_{\rj}$ (Fig. \ref{fig:fig14}b)
and jet strength ${\cal J}_s$ (Fig. \ref{fig:fig14}c) are plotted as a function of $r_{\rj sh}$. The jet shock $r_{\rj sh}$ is formed closer to the BH, if the spin parameter is higher.
However, the most interesting aspect is that the jet shock becomes stronger as $r_{\rj sh}$
shifts to larger values for a BH of same spin. Infact, the compression ratio for $a_s=0.99$
is quite high, and therefore should be a good site for particle acceleration. This nature of
jet shock is completely opposite of accretion shocks, where the accretion shock gets
stronger as it is located closer to the BH (\eg Fig. \ref{fig:fig11}b).
In Fig.(\ref{fig:fig14}d), maximum possible jet terminal speed $v_{\rm m \infty}$ is plotted as a function of
$a_s$. Maximum jet terminal around a BH of $a_s=0.99$ is possible, if the jet originates from a disc that
corresponds to the parameters
of x$_0$ point of the ${\cal E}-\lambda$ space of Fig. (\ref{fig:fig13}B). Therefore, for BH of any spin,
the jet corresponding to its respective x$_0$ point will produce a jet with maximum possible
terminal speed for a BH of that particular spin. Fig. (\ref{fig:fig14}d) is obtained by 
finding the $v_{\rm m \infty}$ for BHs of each $a_s$.
Therefore, BH spin automatically does not produce very high speed jet, but the maximum terminal speed
of a jet definitely increases with increasing BH spin.

\section{Discussions and Summary}\label{sec:conclusn}
We have studied accretion$-$ejection solutions in full general relativistic prescription, where the
fluid is described by variable
adiabatic index ($\Gamma$) EoS around 
Kerr BHs. The mass outflow rate from accretion disc has been estimated self-consistently by solving
accretion equations of motion along the equatorial plane and jet solutions along VZS. 

The accretion solution has two major improvements over some of the previous studies. We consider
a variable $\Gamma$ EoS for electron$-$proton fluid. However, we have not considered heating or cooling
processes in our analysis of the disc, because our main focus is to obtain jets self-consistently
from the disc and how the BH spin affects the jets. Suffice is to say that we have in numerous previous
occasions studied dissipative accretion discs around non-rotating BHs \citep{cd07,lrc11,kc13,kc14,kcm14,ck16,lckhr16}, and based on our accumulated knowledge, we are confident that
the results presented here will qualitatively remain the same, although quantitative changes are not ruled out.
While discussing Figs. (\ref{fig:fig3}a-e), some of the solutions (dotted) related to the
accretion solutions were termed wind type. However,
these solutions are not actual wind or outflow solution, since winds and jets flow in the off-equatorial
direction, while the so-called wind solutions of Figs. (\ref{fig:fig3}a-e) are along the equatorial plane.

One of the reasons to consider accretion discs and jets in full general relativity is that
relativistic formulation of jets is generally faster than the Newtonian versions, since the coupling
between gravity and thermal terms (middle term of ${\cal N}_{\rj}$ in equation \ref{dvdrj.eq}) is missing in the latter.
If one compares equation (\ref{dvdrj.eq}) with the equation of motion of radial or conical outflow
in Schwarzschild metric, then it is easy to identify significance of each of the three terms of
${\cal N}_{\rj}$ (see discussion related to equation \ref{dvdrj.eq}). The comparison also shows that the first two terms of ${\cal N}_j$ for conical flow are positive definite and only the third of the gravity term
is negative. Therefore only one sonic point is possible in conical flow in Schwarzschild metric.
This is also mostly true for rotating matter flowing out along the VZS around BHs of low spin.
However, for higher spin parameter, the metric component along the VZS changes sign from being
highly negative to positive. This causes the jet to become transonic very close to the BH.
And because the same term flips sign along the VZS, this may decelerate the supersonic flow and may cause
the formation of multiple sonic points and shocks in jets.

The jet streamline and cross-section are a major issue in analytical studies of accretion-jet
system.
It has been shown earlier that VZS are the surfaces of constant angular momentum and also of constant entropy for fluid flow above the equatorial plane \citep{c85}.
In addition, numerical simulations showed that the entire PSD
generates bipolar jets
\citep{mrc96,lmc98,dcnm14,lckhr16}. 
So, we choose the foot point ($r_{\rm b}$) of the jet streamline located half way
between the accretion shock and inner critical point of the disc, but on the surface of
PSD. Once we obtain the base
of the streamline or $r_{\rm b}$, we solve for VZS
from the disc properties at $(r_{\rm b}, \theta_{\rm b})$ on PSD surface. Therefore, the VZS is the streamline,
and the cross-sectional area
between $\rci$ and $\rsh$ orthogonal to the VZS is the jet cross-sectional area. This simplifies
the jet structure and make the problem tractable analytically. With these considerations,
the jet streamline 
and the associated cross-section depend on the disc parameters like ${\cal E}$ and $\lambda$,
as well as on the spin of the BH, which is a major improvement on the jet geometry assumed in
the pseudo-Newtonian regime \citep{cd07,kc13,adn15}. Moreover,
the jet solution from the disc is launched with very low velocity along the streamline,
but becomes supersonic at a short distance from the base of the jet. This is because the 
metric term $h_p$ along the VZS for highly spinning BH powers the jet.
This entire analysis was at first
implemented by us in our previous paper for Schwarzschild geometry \citep{ck16};
now we have upgraded it to the Kerr metric.

The jet generated depends on the compression ratio ($R$) of the accretion shock,
the upward thrust of the
PSD and the ratio between jet cross-section with the surface of the PSD. So the mass outflow rate
can be shown to increase with the increasing $R$; however, within few $\rg$ of the BH, the upward thrust and/or the fractional jet cross-section may reduce, making the mass outflow rate
to dip. The relative mass outflow rate for given disc energy parameters also increases with
the spin of the BH. However, the angular momentum range of the accretion discs to obtain such jets
moves to the lower values if the BH spin parameter is increased.
The dependence of the terminal speed of the jet with BH spin is not straightforward, since
the jet is generated by the accretion disc. And accretion disc properties depend
as much on the properties of the central BH as on the boundary conditions of the disc
itself. So if the accretion disc parameters are of moderate values, the jet produced will
be weak, no matter what the value of BH spin parameter is. However, the maximum jet speed for steady
jets increases with the spin of the BH. The maximum jet speed obtained in this paper
for $a_s=0.99$ is about $v_{\rj \infty}\sim 0.35$, which is not truly relativistic.
It is to be remembered that jets from around compact objects like BHs are not always ultrarelativistic.
Not only the terminal speed of jets varies in various objects, even for the same object the jet varies
in strength in different epochs \citep{msa12}.  
Having said that, in this paper we were interested to find out the effect
of BH spin on jet formation and therefore no additional accelerating process for the jet
was considered. Thermally driven jets are unlikely to generate
relativistic terminal speed. Previous studies of accretion-jet system in the pseudo-Newtonian regime
could generate jets
of terminal speed $\sim$few$\times 0.01$ \citep{cd07,kc13}. However, if these jets are accelerated by radiation from
the disc then the terminal speed achieved is $\sim 0.3$ \citep{kcm14}! Even if there is no radiation driving,
then shock oscillation can power jets to around $\sim$few$\times 0.1$ \citep{lckhr16}. In the present paper,
consideration of relativistic equations of motion in Kerr metric, relativistic EoS and obtaining the
VZS from the accretion disc, all have contributed in obtaining  
terminal speeds that are atleast an order of magnitude faster (compared to jets in pseudo-Newtonian regime), 
even for thermally driven jets.
Moreover, in this paper we have only studied the electron$-$proton jets, which are heavier. A finite
proportion of electron$-$positron pair might be conducive for radiative or magnetic acceleration to much higher terminal speeds.
Infact, radiation momentum deposition from luminous discs onto a jet
composed of protons, electrons and positrons can very well accelerate jets to terminal speeds in excess of 90
percent of light speed \citep[e. g., fig. 11 of][]{vkmc15}.   
Therefore, investigation of radiative acceleration of self-consistent jets from accretion discs
around Kerr BHs, in the footsteps of \citet{kcm14}, should yield encouraging results.

In this paper, we studied the jet solutions ejected from the accretion disc. In all our previous studies of
accretion-jet system around a non-rotating BH, we found only monotonic jet solution. In this paper, we find
all possible solutions for jets too. Therefore, not only we have strong jets that become transonic very close to the BH or shocked jets or weak jets that become transonic at large distances, but also bound jet solutions. That means there may be accretion shocks
but there may not be jet if the accretion disc parameters fall in a particular part
of the parameter space. The bound jet solutions may qualify as failed jets.
In this paper, we have mapped various jet solutions in terms of the accretion disc parameters,
which will give an idea about which range of disc parameters will generate jets and which will not. 

The shock in jet is rather strong $>4$, especially around highly spinning BH. While the accretion
shock becomes stronger as it shifts closer to the BH, the shock in jet becomes stronger as it moves
outwards. And jet begins to harbour shocks at $a_s>0.6$. So detection of strong jet shocks close to the 
BH and in the jet might indicate that the BH is spinning. Moreover, the presence of shock in the jet as well as in the accretion disc would produce hot flow around the equatorial plane and also close to the poles too. 
Since the accretion and jet shock are obtained very close to the horizon, therefore if radiation hydrodynamics
of such shocks indeed
produce soft gamma-ray tail, then it may vindicate \citet{lrwbpg11}.

One may also remember that there are few studies of particle
acceleration in accretion disc shocks \citep{lb07,lb16}.
With the presence of strong shocks both in accretion ($R \gsim 3$)
and jets ($R_{\rj}\gsim 4$), it would be worthwhile to study particle acceleration
in such accretion$-$ejection system, especially around highly spinning BH.

\section*{Acknowledgment}%
The authors acknowledge the anonymous referee for helpful suggestions.

{}

\appendix
\section{Calculation of \hp}\label{app:gpp}
The tangent on jet streamline at any point can be expressed as
\begin{equation}
 x_p=mr_{\rm j}{\rm sin}\theta_{\rm j}+c_i,
 \label{tngt.eq}
\end{equation}
where $m=(1-r_{\rj}{\rm tan}\theta_{\rj}\theta_{\rj}^{'})/({\rm tan}\theta_{\rj}+r_{\rj}\theta_{\rj}^{'})$ and $c_i$
are the slope and intercept, respectively. We calculate $\theta_{\rj}={\rm sin}^{-1}
\left(\left[-s-\sqrt{s^2-4Z_{\phi}^2a_{\rm s}^2}\right]/2a_{\rm s}^2\right)^{1/2}$ and 
$\theta_{\rj}^{'}=d\theta_{\rj}/dr_{\rj}=[s^{\prime} {\rm tan}\theta_{\rj}/2]/\sqrt{s^2-4Z_{\phi}^2a_{\rm s}^2}$,
from equation (\ref{vZp.eq}).  
We define $s=-[a_{\rm s}^2\lambda_{\rj}Z_{\phi}^2-2a_{\rm s}r_{\rj}Z_{\phi}^2+\lambda_{\rj}(r_{\rj}^2+a_{\rm s}^2)^2-
2\lambda_{\rj}^2a_{\rm s}r_{\rj}]/(\Delta_{\rj}\lambda_{\rj})$ and 
$s^{\prime}=ds/dr_{\rj}=[-\lambda_{\rj}s\Delta_{\rj}^{\prime}+2a_{\rm s}(Z_{\phi}^2+\lambda_{\rj}^2)-4\lambda_{\rj}r_{\rj}(r_{\rj}^2+
a_{\rm s}^2)]/(\Delta_{\rj}\lambda_{\rj})$. 
The basis vector along streamline is defined as
\begin{equation}
{\bf e}_p=(\frac{\partial r_j}{\partial x_p}){\bf e}_r+(\frac{\partial \theta_j}{\partial x_p}){\bf e}_\theta,
\label{bsvec.eq}
\end{equation}
where ${\bf e}_p=h_p\hat{e}_p, {\bf e}_r=h_r\hat{e}_r$ and ${\bf e}_\theta=h_\theta\hat{e}_\theta$. Here, $\hat{e}_p, \hat{e}_r$ and 
$\hat{e}_\theta$ are unit basis vectors along the tangent, radial and polar direction at a point, respectively. 
The magnitude of the basis vector is written as
\begin{equation}
 h_p^2=h_r^2\left(\frac{\partial r_{\rm j}}{\partial x_p}\right)^2+h_\theta^2\left(\frac{\partial\theta_{\rm j}}{\partial x_p}\right)^2,
 \label{hp1.eq}
\end{equation}
where $h_r^2=g_{rr}=\Sigma_{\rj}/\Delta_{\rj}$ and $h_\theta^2=g_{\theta\theta}=\Sigma_{\rj}$ are Kerr metric components. 
For the calculation of $h_p$, 
we have taken partial differentiation of equations (\ref{vZp.eq}) and (\ref{tngt.eq}) with respect to $x_p$; we get
\begin{eqnarray}\nonumber
 K^2\left(\frac{\partial r_{\rm j}}{\partial x_p}\right)^2=
 {\rm cot}^2\theta_{\rm j}\left(\frac{\partial\theta_{\rm j}}
 {\partial x_p}\right)^2 \\~~\mbox{and}~~ {\rm cos}^2\theta_{\rm j}=C_r^2\left(\frac{\partial r_{\rm j}}
 {\partial x_p}\right)^2+
 C_{\theta}^2\left(\frac{\partial\theta_{\rm j}}{\partial x_p}\right)^2,
 \label{bsisdv.eq}
\end{eqnarray}
where $C_r=[{\rm cos}^2\theta_{\rj}-r_{\rj}^2K^{\prime}-r_{\rj}K(2+r_{\rj}K){\rm sin}^2\theta_{\rj}]/(1+r_{\rj}K)^2$, 
$C_{\theta}=r_{\rj}{\rm tan}\theta_{\rj}[{\rm cos}^2\theta_{\rj}+r_{\rj}K/(1+r_{\rj}K)]$, 
 $K={s^{\prime }}/{2(s^2-4a_{\rs}^2Z_{\phi}^2)^{1/2}}$, 
 $K^{\prime}=dK/dr_{\rj}=K(s^{\prime\prime}-4sK^2)/s^{\prime}$ and 
$s^{\prime\prime}=-2[s+s^{\prime}\Delta_{\rj}^{\prime}+2(3r_{\rj}^2+a_{\rs}^2)]/\Delta_{\rj}$ .
Using equation (\ref{bsisdv.eq}) in equation (\ref{hp1.eq}), then we get the expression of $h_p$, 
\begin{equation}
h_p^2=\frac{{\rm cos}^2\theta_{\rj}[h_r^2+h_{\theta}^2K^2{\rm tan}^2\theta_{\rj}]}{[C_r^2+C_{\theta}^2K^2{\rm tan}^2\theta_{\rj}]}
 \label{hp.eq}
\end{equation}

\end{document}